\documentclass[11pt]{article}
\usepackage{jheppub}

\usepackage{color}
\usepackage{colordvi}
\usepackage{changes}

\usepackage{epsfig,epsf}
\usepackage{amsmath}
\usepackage{amsthm}
\usepackage{amsfonts}
\usepackage{amssymb}
\usepackage{dsfont}
\usepackage[mathscr]{euscript}
\usepackage{eufrak}
\usepackage{epstopdf}
\usepackage{multirow}
\usepackage{mathrsfs}

\usepackage{rotating}

\usepackage{marvosym}

\usepackage{todonotes}

\usepackage{subcaption}
\usepackage{array}   
\newcolumntype{C}{>{$}c<{$}}

\usepackage{slashed}
\usepackage{cancel}

\usepackage[active]{srcltx}

\newcommand{\tH}{\widetilde{H}}
\newcommand{\tE}{\widetilde{E}}

\newcommand \widebar [1] {\overline{#1}}

\def\II{\hbox{{1}\kern-.25em\hbox{l}}}

\def\II{\hbox{{1}\kern-.25em\hbox{l}}}

\makeatletter
\renewcommand\@fpheader{}
\renewcommand\@journal{}
\makeatother

\title{
Finite-$t$ and target mass corrections for the short-distance expansion
of quasi(pseudo) GPDs}

\author[a]{Vladimir M. Braun,}
\author[b]{and Hua-Yu Jiang}

\affiliation[a]{
   Institut f\"ur Theoretische Physik, Universit\"at
   Regensburg, D-93040 Regensburg, Germany}
\affiliation[b]{
Institute of Particle and Nuclear Physics, Henan Normal University, Xinxiang 453007, Henan, P. R. China}

\emailAdd{vladimir.braun@ur.de}

\emailAdd{jianghuayu@htu.edu.cn}

\abstract{
We calculate the ``kinematic'' corrections $t/P_z^2$ and $m_N^2/P_z^2$ to the short distance 
expansion of gauge-invariant nonlocal quark-antiquark operators sandwiched 
between nucleon states with different momenta. Here $t$ is the momentum 
transfer, $m_N$ is the nucleon mass and $P_z$ is the momentum component
in the direction of the quark-antiquark separation, which is assumed to be large.
These matrix elements can be calculated in lattice QCD  and, at leading twist, 
expressed in terms of moments of the generalized parton distrubutions (GPDs).
Our results allow one to control one of principal uncertainties in such calculations and extend 
their region of applicability to larger momentum transfers, 
which is important in the quest to access the three-dimension image of the proton. 
The calculated corrections turn out to be significant for a realistic lattice QCD setup. 
       }


\begin{document}
\maketitle


                          \section{Introduction}\label{sec:intro}


Generalized parton distributions
(GPDs)~\cite{Muller:1994ses,Ji:1996nm,Radyushkin:1997ki} encode the information on the transverse position of quarks and gluons in
the proton in dependence on their longitudinal momentum.  This is  an active research topic and a major science goal for the
12 GeV Upgrade at Jefferson Lab \cite{Dudek:2012vr} and
planned new accelerators in U.S. \cite{AbdulKhalek:2021gbh,Accardi:2023chb} and China \cite{Anderle:2021wcy}. 

The main challenge of all GPD studies is that the quantities of interest are functions of three kinematic variables.
Their extraction requires a massive amount of data and very high precision for both experimental and theory inputs.
The future GPD determinations will therefore have to be based on global fits of all available experiments,
the constraints from lattice measurements, and parton distribution functions (PDFs) in the forward limit.

In particular the importance of lattice input is gaining attention, see e.g. \cite{Burkert:2025gzu}, 
and the techniques to perform lattice calculations of GPDs and/or GPD moments 
have been developing rapidly in recent years
\cite{Chen:2019lcm,Alexandrou:2020zbe,Lin:2020rxa,CSSMQCDSFUKQCD:2021lkf,Alexandrou:2021bbo,Bhattacharya:2022aob,%
Bhattacharya:2023ays,Bhattacharya:2023jsc,Bhattacharya:2024qpp,Bhattacharya:2024wtg,HadStruc:2024rix,Schoenleber:2024auy,%
Bhattacharya:2025yba,Chu:2025kew,Gao:2025inf}. 
One perceived advantage of lattice approach is that the zero skewedness limit $\xi =0$ can be accessed directly, 
at least in principle, whereas the extraction from experimental data on DVCS requires a nontrivial extrapolation.

In this work we address one particular issue in lattice GPD calculations, taking into account power-suppressed (for large momenta)
contributions to the nonlocal equal-time  quark-antiquark correlation functions 
$\langle p'|\bar q(z)\Gamma q(0)|p\rangle$  
of the type $m_N^2/P_z^2$ and $t/P_z^2$ where $t=(p'-p)^2$ is the momentum transfer, $m_N$ is the nucleon mass and $P_z$ is the 
target momentum component in the direction of the quark-antiquark separation, which is assumed to be large.  
(Precise definitions will be given below.) 
Such contributions are usually referred to as ``kinematic power corrections'' because they do not involve new 
nonperturbative inputs in addition to the leading twist GPDs.
They can be large and have to be taken into account. Indeed, the
transverse spatial position of partons in the target is Fourier
conjugate to the momentum transfer in the scattering process.
Hence the resolving power of the transverse distance encoded in the GPDs is 
directly limited by the range of the invariant moment transfer $t$ which can be used
in their extraction. Theoretical control over power corrections
$\sim (t/P^2_z)$  is therefore crucial for three-dimensional imaging.

The similar corrections also impact GPD extraction from DVCS. They were derived using two different techniques
\cite{Braun:2011zr,Braun:2011dg,Braun:2012bg,Braun:2012hq,Braun:2014sta,Braun:2020zjm,Braun:2022qly,Braun:2025xlp,%
Martinez-Fernandez:2025gub,Martinez-Fernandez:2025jvk,Martinez-Fernandez:2025rcg}
so that detailed results are available to twist-6 accuracy.
A typical size of kinematic corrections to DVCS for $|t|/Q^2\le 1/4$ was found to be of the order of 10\% for asymmetries, 
but they could be as large as 100\% for the cross section in certain kinematics. 

In this work we use the approach developed in  Ref.~\cite{Braun:2011dg} to calculate kinematic contributions
to the position-space quark-antiquark correlation functions $\langle p'|\bar q(z)\Gamma q(0)|p\rangle$
which give rise to both, quasi- and pseudo-GPDs, after a Fourier transform. To the tree-level accuracy that 
is relevant for this work, 
there is no difference. The structure of the twist-four kinematic corrections  proves to be more complicated 
as compared to the DVCS case, which is due to the fact that  quark-antiquark correlators off light-cone do not have 
good symmetry properties under conformal transformations. As a consequence, 
the short-distance expansion of the vector and axial-vector correlation functions in terms of the GPD moments can be 
obtained, but we have not been able to find compact analytic expressions in terms of GPDs themselves.    
Note that the kinematic power corrections depend nontrivially on the quark positions.   
The choice with the quark field at the origin is advantageous for lattice calculations \cite{Bhattacharya:2022aob} 
and also leads to somewhat simpler expressions for power corrections, so that we concentrate on this case.  
For the zero skewedness limit $\xi =0$ the results are simplified considerably. However, kinematic power corrections 
for the correlation functions at $\xi =0$ also involve the higher-order terms in the expansion of the 
GPD moments over the powers of the skewedness parameter, albeit with a small coefficient. 

The presentation is organized as follows.
Sect.~\ref{sec:general} is introductory, it contains
general definitions and specifies our notation and conventions.
In Sect.~\ref{sec:framework} we start with a simple example and explain the general framework and the procedure for the calculation
of kinematic power corrections using the formalism of Ref.~\cite{Braun:2011dg}, see also appendix~\ref{app:A}. 
For definiteness, in this section the vector 
operator is considered. Detailed results for this case are presented in Sect.~\ref{sec:results} 
and also in appendix~\ref{app:B} for the first few moments..
The axial-vector operator case is very similar; the results are summarized in appendix~\ref{app:C}.    
Numerical estimates of the size of the kinematic corrections 
for a realistic GPD model are presented in Sect.~\ref{sec:MP}. 
The final Sect.~\ref{sec:conclusions} is reserved for a summary and conclusions.


               \section{Kinematics, notation and conventions}\label{sec:general}


               \subsection{Generalized Parton Distributions }\label{sec:defGPD}

Generalized Parton Distributions (GPDs) are defined by matrix elements of the light-ray operators
\begin{align}
O^{(\slashed n)}(z_1n,z_2n) &= \bar q(z_1n) \slashed{n} [z_1n,z_2n] q(z_2n)\,,
\notag\\
O^{(\slashed n\gamma_5)}(z_1n,z_2n) &= \bar q(z_1n) \slashed{n}\gamma_5 [z_1n,z_2n] q (z_2n)\,,
\label{LRO}
\end{align}
where $q(x)$ is the quark field of a given flavor, $n^\mu$ is a light-like vector, $n^2=0$,  
$[z_1n,z_2n]$ is the Wilson line connecting the fields, and $z_1$, $z_2$ are real numbers.   
\footnote{We use Minkowski metric, $g_{\mu\nu} = \text{diag}(1,-1,-1,-1)$, 
the antisymmetric tensor with $\epsilon_{0123} = 1$ and $\gamma_5 = i\gamma^0\gamma^1\gamma^2\gamma^3$.} 

For the nucleon target \cite{Diehl:2003ny,Belitsky:2005qn}
\begin{align}\label{defNucleonGPD}
&\langle p^\prime|\mathcal {O}^{(\slashed{n})}(z_1n,z_2n)|p\rangle =
\notag\\&= 
\int_{-1}^1\!\! dx\, e^{-i (Pn)[z_1(\xi-x)+z_2(x+\xi)]}
\biggl\{ \bar u(p') \slashed{n} u(p)  H(x,\xi,t) + \bar u(p')\frac{i\sigma^{n\alpha}\Delta_\alpha}{2m_N} u(p) E(x,\xi,t)
\biggr\}  
\notag\\&=
\int_{-1}^1\!\! dx\, e^{-i (Pn)[z_1(\xi-x)+z_2(x+\xi)]}
\biggl\{ \bar u(p') \slashed{n} u(p) M (x,\xi,t) - \frac{2(Pn)}{2m_N} \bar u(p') u(p) E(x,\xi,t)
\biggr\},  
\notag\\
&\langle p^\prime|\mathcal {O}^{(\slashed{n}\gamma_5)}(z_1n,z_2n)|p\rangle =
\notag\\&= 
\int_{-1}^1\!\! dx\, e^{-i (Pn)[z_1(\xi-x)+z_2(x+\xi)]}
\biggl\{ \bar u(p') \slashed{n}\gamma_5 u(p)  \widetilde H(x,\xi,t) + \frac{(\Delta n)}{2m_N}  \bar u(p')\gamma_5 u(p) \widetilde E(x,\xi,t)
\biggr\},  
\end{align}
where
\begin{align}
M(x,\xi,t) &= H(x,\xi,t) + E(x,\xi,t)\,. 
\end{align}  
Here
\begin{align}
 P_\mu = \frac12 (p+p')_\mu\,, \qquad \Delta_\mu = (p'-p)_\mu\,, \qquad t = \Delta^2,
\end{align}
and the asymmetry parameter $\xi$ is defined as
\begin{align}
\xi = -\frac{(\Delta n)}{2(Pn)}\,.
\label{xi}
\end{align}
Gegenbauer moments of the GPDs are given by reduced matrix elements of local conformal
operators. E.g. for the vector operators
\begin{align}
{O}_N^{(\slashed{n})}(y) &= 
  (-\partial_+)^N \bar q(y) C_N^{3/2}
\left(\frac{\stackrel{\rightarrow}{D}_+-\stackrel{\leftarrow}{D}_+}
           {\stackrel{\rightarrow}{D}_++\stackrel{\leftarrow}{D}_+} \right)\slashed{n}q(y)\,,
\label{ON}
\end{align}
where  $D_\mu = \partial_\mu -i g A_\mu$ is the covariant derivative and
$\partial_+ = \partial_\mu n^\mu$. Let 
\begin{align}\label{iO}
(-i)^N\langle p'|{O}^{(\slashed{n})}_N |p\rangle&=
\bar u(p')\slashed{n}u(p)\, M_{N}^{({n})}(t) - \frac{Pn}{m_N}\bar u(p')u(p)\,E_{N}^{({n})}(t)\,
\end{align}
with
\begin{align}
M^{({n})}_{N}(t) &= \sum_{m,\text{even}}^NM_{Nk}(t)(\Delta n)^m (P n)^{N-m}\,,
\qquad E_{N}^{({n})}(t) = \sum_{m,\text{even}}^{N+1}E_{Nm}(t)(\Delta n)^m (P n)^{N-m}
\label{moexp}
\end{align}  
and similar for the axial vectors,
\begin{align}
\widetilde{H}^{({n})}_{N}(t) &= \sum_{m,\text{even}}^N\widetilde H_{Nk}(t)(\Delta n)^m (P n)^{N-m}\,,
\qquad \widetilde E_{N}^{({n})}(t) = \sum_{m,\text{even}}^{N}\widetilde E_{Nm}(t)(\Delta n)^m (P n)^{N-m}.
\label{Axialmoexp}
\end{align}  
One obtains after a short calculation 
\begin{eqnarray}
\int_{-1}^{1}\!  dx \, C_N^{3/2}\left(\frac{x}{\xi}\right)\,M(x,\xi,t)
&=&
\sum_{m,\text{even}}^N (-1)^m (2\xi)^{m-N} M_{Nm}(t) \,, 
\notag\\
\int_{-1}^{1}\! dx \, C_N^{3/2}\left(\frac{x}{\xi}\right)\,E(x,\xi,t)
&=&
\sum_{m,\text{even}}^{N+1}(-1)^{m}(2\xi)^{m-N} E_{Nk}(t)
\end{eqnarray}
and
\begin{eqnarray}
\int_{-1}^{1}\! dx \, C_N^{3/2}\left(\frac{x}\xi\right)\,\widetilde{H}(x,\xi,t)
&=&
\sum_{m,\text{even}}^N (-1)^{m}(2\xi)^{m-N} \widetilde{H}_{Nm}(t)\,, 
\notag\\
\int_{-1}^{1}\! dx \, C_N^{3/2}\left(\frac{x}\xi\right)\,\widetilde{E}(x,\xi,t)
&=&
\sum_{m,\text{even}}^{N}(-1)^{m}(2\xi)^{m-N} \widetilde{E}_{Nm}(t)\,.
\end{eqnarray}
Note the the sum extends to $N+1$ for $E$ (the D-term) whereas for the other GPDs this contribution is absent~\cite{Diehl:2003ny}.

              \subsection{Quasi(Pseudo)  Distributions }\label{sec:defqGPD}

In the lattice QCD approach built upon the LAMET paradigm \cite{Ji:2013dva}
one considers nonlocal quark-antiquark operators at space-like separations
\begin{align}
Q^{(\gamma_\mu)}(z_1v,z_2v) &= \bar q(z_1v) \gamma_\mu [z_1v,z_2v] q(z_2v)\,,
\label{qLROvector}
\\
Q^{(\gamma_\mu\gamma_5)}(z_1v,z_2v) &= \bar q(z_1v) \gamma_\mu\gamma_5 [z_1v,z_2v] q (z_2v)\,,
\label{qLROaxial}
\end{align}
where $v^2 <0$. We will keep the direction and the normalization of $v^\mu$ arbitrary to maintain explicit 
Lorentz covariance; the usual choice  is $v^\mu = (0,0,0,1)$.

Nucleon matrix elements of the operators \eqref{qLROvector}, \eqref{qLROaxial} for large target momenta can be matched to the matrix elements
of the corresponding light-ray operators \eqref{LRO}, schematically 
\begin{align}
\langle p'|Q^{(\Gamma)}(z_1v,z_2v)| p\rangle = C_\Gamma(z_1v,z_2v;\mu_F^2)\otimes \langle p'|Q^{(\slashed{n})}(z_1n,z_2n)| p\rangle^{\mu_F^2}  
+ \mathcal{O}(1/(Pv))\,,
\label{LAMET}
\end{align} 
where $\mu_F$ is the factorization scale. This matching is well understood and the coefficient functions $C_\Gamma$ have been 
calculated to the NLO accuracy~\cite{Ji:2015qla,Xiong:2015nua,Liu:2019urm,Radyushkin:2019owq}. 
Information on the GPDs can be harvested from these  correlation functions 
in several different ways, either directly in position space, or via a Fourier transform leading to quasi- (qGPD)  or pseudo- (pGPD)
distributions.
The subject of this paper are ``kinematic'' power-suppressed corrections to the factorization theorem in \eqref{LAMET}. 
They will be considered at tree level and to this accuracy the difference between pGPDs and qGPDs is irrelevant.  

Power corrections to the matrix element of the vector operator \eqref{qLROvector} receive contributions of both vector 
and axial-vector GPDs --- $M,E$ and $\widetilde H,\widetilde E$, respectively. It is convenient to separate them,
\begin{align}
  \langle p'|Q^{(\gamma_\mu)}|p\rangle &= \langle p'|Q^{(\gamma_\mu)}|p\rangle^{V} 
+ \langle p'|Q^{(\gamma_\mu)}|p\rangle^{A}.
\end{align}  
For vector contributions, 
we will employ the following decomposition of the nucleon matrix element in invariant amplitudes: 
\begin{align}
 \langle p'|Q^{(\gamma_\mu)}|p\rangle^V &=   
 \bar u(p')\slashed{v} u(p)
\biggl\{\frac{v^\mu}{v^2} \,\mathcal{M}^{(v)}
+  \left(\frac{P^\mu}{(Pv)} -  \frac{v^\mu}{v^2}\right)
\mathcal{M}^{(P)}
+ \left(\frac{\Delta^\mu_\perp}{(Pv)}\right) 
\mathcal{M}^{(\Delta)}\biggr\}
\notag\\&\quad
+ \bar u(p')\left(\gamma_\mu- \frac{P_\mu\slashed{v}}{(Pv)}\right) u(p) \mathcal{M}^{(\gamma)} 
\notag\\&\quad
  - \frac{(Pv)}{m_N}\bar u(p')u(p)
\biggl\{ \frac{v^\mu}{v^2} \, \mathcal{E}^{(v)}
+  \left(\frac{P^\mu}{(Pv)} - \frac{v^\mu}{v^2}\right) \mathcal{E}^{(P)} 
+ \left(\frac{\Delta^\mu_\perp}{(Pv)}\right) 
\mathcal{E}^{(\Delta)}\biggr\},
\label{Vstructures}
\end{align}
For the axial-vector contributions which appear at twist-three level, we have chosen
to leave the results in the form how they come out naturally in the calculation
\begin{align}
 \langle p'|Q^{(\gamma_\mu)}|p\rangle^A &=   
 \bar u(p')\left(\gamma_\mu- \frac{P_\mu\slashed{v}}{(Pv)}\right) u(p) \widetilde{\mathcal{H}}^{(\gamma)} 
+ \frac{\widetilde{\Delta}^\perp_\mu}{(Pv)} \bar u(p')\slashed{v}\gamma_5u(p) \widetilde{\mathcal{H}}^{(\tilde\Delta)}
\notag\\&\quad
+  \frac{\widetilde{\Delta}^\perp_\mu}{(Pv)}  \frac{(\Delta v)}{2 m_N}\bar u(p')\gamma_5u(p)  \widetilde{\mathcal{E}}^{(\tilde \Delta)},   
\label{Astructures}
\end{align}
where
\begin{align}
\Delta_\perp^\mu = \Delta^\mu - P^\mu \frac{(\Delta v)}{(Pv)}\,,\qquad
\epsilon^\perp_{\mu\nu} = \epsilon_{\mu\nu\alpha\beta} \frac{P_\alpha v_\beta}{(Pv)}\,, \qquad 
 \widetilde{\Delta}^\perp_\mu =  i \epsilon^\perp_{\mu\nu}  \Delta^\nu.
\end{align}
The last two contributions in \eqref{Astructures} can be rewritten in terms of the seven Lorentz/Dirac structures 
defined in \eqref{Vstructures} and the single new structure $\bar u(p') \sigma_{\mu\nu}v^\nu u(p)$ 
using that
\begin{align}
i\epsilon^{\rho\nu\sigma\mu} v_\rho \Delta_\nu  \bar u (p') \gamma^\sigma\gamma_5 u(p) 
&=  2 \bar u(p') \big[ (Pv) \gamma_\mu -  P_\mu \slashed{v} \big]u(p)\,,
\notag\\
  i\epsilon^{\rho\nu\sigma\mu} v_\rho P_\nu  \bar u(p') \gamma^\sigma\gamma_5 u(p) 
&= \frac12 \bar u(p')[ (\Delta v)\gamma_\mu  -  \Delta_\mu \slashed{v}]u(p)   
- m_N \bar u(p')[i \sigma_{\mu\xi} v^\xi]u(p)\,,   
\notag\\
 i\epsilon^{\rho\nu\sigma\mu} \Delta_\rho P_\nu \bar u(p') \gamma^\sigma\gamma_5 u(p)
& =  2 m_N P_\mu \bar u(p')  u(p)  - 2 \left(m_N^2- \frac{t}{4}\right)  \bar u(p') \gamma_\mu u(p)\,. 
\end{align}
Thus, it total, there are eight independent Lorentz-invariant amplitudes, in agreement with the counting in 
Ref.~\cite{Bhattacharya:2022aob}.
We prefer to leave the results in this form as it leads to simpler expressions.

Each invariant amplitude can be written as a sum of contributions of increasing twist, e.g.,
\begin{align}
\mathcal{M}^{(P)} &= \mathcal{M}^{(P)}_{t2} + \mathcal{M}^{(P)}_{t3} + \mathcal{M}^{(P)}_{t4} +\ldots
\end{align}
It is easy to verify that $\mathcal{M}^{(v)}$,  $\mathcal{M}^{(P)}$,  $\mathcal{E}^{(v)}$ and  $\mathcal{E}^{(P)}$
amplitudes start at leading power and differ by corrections, 
$\mathcal{M}^{(v)}- \mathcal{M}^{(P)} = \mathcal{O}(1/(Pv)^2)$, $\mathcal{E}^{(v)}- \mathcal{E}^{(P)} = \mathcal{O}(1/(Pv)^2)$. 
Our goal will be to calculate kinematic power corrections to these
amplitudes to the $1/(Pv)^2$ accuracy. All other amplitudes start at twist three
so that the power corrections for them, terms $\sim t/(Pv)^2, m_N^2/(Pv)^2$, involve contributions of twist-five operators 
and are beyond our accuracy.   

It should be noted that the dependence of power corrections on quark positions $z_1,z_2$ is nontrivial, and kinematic 
power corrections is fact restore translation invariance of the invariant amplitudes which is broken by the twist
expansion \cite{Braun:2012bg,Braun:2014sta,Braun:2023alc}. 
The choice with the quark field at the origin is advantageous for lattice calculations \cite{Bhattacharya:2022aob} 
and it also leads to somewhat simpler expressions for power corrections, so that we concentrate on this case,
$z_1=z$, $z_2=0$.    

The complete expression for the kinematic twist-three contributions is available from \cite{Belitsky:2000vx,Kivel:2000rb,Braun:2023alc}
in nonlocal form in terms of the GPDs. The twist-four contributions turn out to be are more complicated, 
and it is unlikely that a compact representation of this type can be found. In this paper we present all results in the 
form of a short-distance expansion, e.g. 
\begin{align}
 \mathcal{M}^{(P)} = \sum_{N}\frac{\Gamma(3/2)[iz(Pv)]^N}{4^N\Gamma(N+3/2)}\,  \mathcal{M}^{(P)}_N\,,
\label{SDE}
\end{align}
and similar for the other amplitudes. The (dimensionless) variable $\tau = z(Pv)$ is often referred to as Ioffe 
time~\cite{Gribov:1965hf,Ioffe:1969kf,Braun:1994jq}. 


               \section{Operator Product expansion in off-forward kinematics}\label{sec:framework}


               \subsection{Simple example}\label{sec:example}

In this section we explain the origin and structure of the kinematic power corrections on a simple example. 
Let $\mathcal{O}_{\{\mu_1\ldots\mu_{N+1}\}}$ be the leading twist operators involving Gegenbauer polynomials in covariant derivatives.
E.g. 
\begin{align}
\mathcal{O}_{\{\alpha\beta\gamma\}} =\mathrm{Sym}_{\alpha\beta\gamma}\left[
\frac{15}{2}\bar q \gamma_\alpha\!\stackrel{\leftrightarrow}{D}_{\beta}
\stackrel{\leftrightarrow}{D}_{\gamma}  q
-\frac{3}{2} \partial_{\beta} \partial_{\gamma}\bar q \gamma_\alpha
q \right]  - \mathrm{traces}\,.
\end{align}
Here and below $\{\ldots\}$ 
denotes the symmetrization of all enclosed Lorentz indices and the subtraction of traces.

Picking up the $N=2$ contribution in the OPE for the nonlocal operator $Q^{(\slashed v)}$  
\eqref{qLROvector}%
\footnote{We take the $\slashed v$ projection for this example because it does not contain twist-three
contributions.}
one obtains
\begin{align}
 \langle \bar q (z v) \slashed{v} q (0) \rangle_{N=2} &= 
v^\alpha v^\beta v^\gamma \biggl\{\frac{z^2}{60} \langle \mathcal O_{\{\alpha\beta\gamma\}}\rangle + 
   \frac{z^2}{12} \langle\partial_{\{\alpha} O_{\beta\gamma\}}\rangle +  \frac{3 z^2}{20} \langle \partial_{\{\alpha}\partial_\beta O_{\gamma\}}\rangle\biggr\}\notag\\&\quad
+ v^2 v_\alpha\frac{z^2}{36} \langle \partial^\mu\mathcal{O}_{\mu\alpha}\rangle + \ldots\,, 
\label{N=2}
\end{align}
where we use $\langle\ldots\rangle$ as a shorthand notation for the matrix element $\langle p'|\ldots|p\rangle$.
Ellipses stand for the contribution of the ``genuine'' twist-four operator $\bar q g \widetilde F^{\mu\nu}\gamma_\mu\gamma_5q$~\cite{Shuryak:1981kj}. 
Its matrix element defines the lowest moment of the twist-four GPDs and, by definition, is unrelated to kinematic power corrections.
The tree-level coefficients of the leading-twist operators in the first line of Eq.~\eqref{N=2} can be calculated trivially, using 
explicit expressions for the Gegenbauer polynomials. Conversely, as explained below,  the coefficient $1/36$ of the twist-four operator 
in the second line is nontrivial.

Consider a scalar target and define the expansion of the corresponding GPD $H(x,\xi,t)$ in terms of conformal moments 
similar to Eqs.~\eqref{iO}, \eqref{moexp}%
\footnote{For simplicity we do not include a $D$-term.}:
\begin{align}
(-i)^N\langle p'|{O}^{(\slashed{n})}_N |p\rangle &= (Pn)  \sum_{m,\text{even}}^{N}H_{Nm}(t)(\Delta n)^m (P n)^{N-m}. 
\label{eq:scalarME}
\end{align}
Then
\begin{align}
\langle \mathcal O_{\{\alpha\beta\gamma\}}\rangle &=
- H_{20}\biggl[P_\alpha P_\beta P_\gamma 
- \frac{P^2}{6}\Big( g_{\alpha\beta} P_\gamma +  g_{\alpha\gamma} P_\beta  + g_{\gamma\beta} P_\alpha\Big)\biggr]
\notag\\&
-  \frac13 H_{22} \biggl[
 \Big(\Delta_\alpha \Delta_\beta P_\gamma + \Delta_\alpha \Delta_\gamma P_\beta + \Delta_\gamma \Delta_\beta P_\alpha\Big)
- \frac{\Delta^2}{6} \Big( g_{\alpha\beta} P_\gamma +  g_{\alpha\gamma} P_\beta  + g_{\gamma\beta} P_\alpha\Big)
\biggr],
\notag\\
\langle \partial_{\{\alpha} \mathcal O_{\beta\gamma\}}\rangle &=
- \frac13 H_{10} \biggl[
 \Big(\Delta_\alpha P_\beta P_\gamma + \Delta_\gamma  P_\alpha P_\beta + \Delta_\beta P_\alpha P_\gamma\Big)
- \frac{P^2}{6} \Big( g_{\alpha\beta} \Delta_\gamma +  g_{\alpha\gamma} \Delta_\beta  + g_{\gamma\beta} \Delta_\alpha\Big)
\biggr],
\notag\\
\langle \partial_{\{\alpha} \partial_\beta \mathcal O_{\gamma\}}\rangle &=
- \frac13 H_{00} \biggl[
 \Big(\Delta_\alpha \Delta_\beta P_\gamma + \Delta_\alpha \Delta_\gamma P_\beta + \Delta_\gamma \Delta_\beta P_\alpha\Big)
- \frac{\Delta^2}{6} \Big( g_{\alpha\beta} P_\gamma +  g_{\alpha\gamma} P_\beta  + g_{\gamma\beta} P_\alpha\Big)\biggr],
\notag\\
\langle \partial^{\mu} \mathcal O_{\{\mu\alpha\}}\rangle &= i^2 \Delta^\mu H_{10} \biggl[P_\mu P_\alpha -\frac14 g_{\mu\alpha} P^2\biggr]
= \frac14 H_{10} \Delta_\alpha P^2 .
\end{align}
Here
\begin{align}
 P^2 = M^2 -\frac{\Delta^2}{4}\,,
\label{P2}
\end{align}
where $M$ is the target mass, and we do not show the $t$-dependence of the moments $H_{Nm}(t)$ for brevity.
 Using these expressions one immediately obtains
\begin{align}
v^\alpha v^\beta v^\gamma \langle O_{\{\alpha\beta\gamma\}}\rangle&=
- H_{20} (Pv)^3 \biggl[ 1 -  \frac12 \frac{P^2 v^2}{(Pv)^2}\biggr]
- H_{22} (Pv) (\Delta v)^2  \biggl[1 - \frac16 \frac{\Delta^2 v^2}{(\Delta v)^2}\biggr] 
\notag\\&=
- (Pv)^3\biggl[
H_{20} + 4\eta^2 H_{22}  - \frac12  \frac{P^2 v^2}{(Pv)^2}H_{20} - \frac16 \frac{\Delta^2 v^2}{(P v)^2}H_{22}\biggr], 
\notag\\
v^\alpha v^\beta v^\gamma \langle \partial_{\{\alpha} O_{\beta\gamma\}}\rangle&=
- (\Delta v)(Pv)^2 H_{10} \Big[1- \frac16 \frac{P^2 v^2}{(P v)^2}\Big]
= 2 \eta H_{10} (Pv)^3  \Big[1- \frac16 \frac{P^2 v^2}{(P v)^2}\Big],
\notag\\
v^\alpha v^\beta v^\gamma \langle \partial_{\{\alpha} \partial_\beta O_{\gamma\}}\rangle&=
-(\Delta v)^2(Pv) H_{00} \Big[1-  \frac16 \frac{\Delta^2 v^2}{(\Delta v)^2}\Big]
= - (Pv)^3H_{00} \Big[4\eta^2 - \frac16 \frac{\Delta^2 v^2}{(P v)^2}\Big],
\notag\\
v^2 v^\alpha \langle \partial^{\mu} \mathcal O_{\{\mu\alpha\}}\rangle &= 
\frac14 H_{10} v^2 (\Delta v)  P^2 = - \frac12 \eta H_{10} (Pv)^3 \frac{P^2v^2}{(Pv)^2},  
\end{align}
where 
\begin{align}
 \eta &= -\frac{(\Delta v)}{2(P v)}
\label{eta}
\end{align}
is the ``lattice asymmetry parameter'' defined with respect to the space-like projection $v^\mu$ in contrast to the usual definition 
\eqref{xi} using a light-like vector.

Collecting everything, we obtain
\begin{align}
 \langle p'|\bar q(z v) \slashed{v} q (0)|p\rangle_{N=2} &=
\frac{(iz)^2}{60} (Pv)^3
\biggl\{ 
H_{20} + 4\eta^2 H_{22} -10\eta H_{10} + 36 \eta^2 H_{00}
\notag\\&\quad
+ \frac16 \frac{v^2}{(P v)^2} \Big[
{ - 3 P^2 H_{20}} 
{-  \Delta^2 H_{22}} 
{+ 10 P^2 \eta H_{10}} 
{- 9  \Delta^2 H_{00}} 
\Big]
\notag\\&\quad
+\frac56
 \eta H_{10}  \frac{P^2 v^2}{(P v)^2}
\biggr\}.
\end{align}
Here the first line displays the leading-twist expression. The second line contains power 
corrections arising from the trace subtraction in the leading twist operators; these corrections represent
a straightforward generalization of the Nachtmann corrections to the deep-inelastic scattering \cite{Nachtmann:1973mr}.
Finally, the contribution from the divergence of the leading twist operator in the last line is unique for off-forward reactions. 

We observe that, first, the zero lattice asymmetry limit $\eta =0$ does not remove the sensitivity to higher-order coefficients 
in the $\xi^2$ expansion of the GPD \eqref{eq:scalarME} completely; the $H_{22}$ term remains to be present in the power correction, albeit with a 
small coefficient. Second, the contribution of the twist-four operator $\partial^{\mu} \mathcal O_{\{\mu\alpha\}}$ includes the factor $\eta$
and vanishes in the zero asymmetry limit. Both properties hold also for higher $N$-moments and for a spin-1/2 target.

Finally, we explain why calculating the kinematic contributions from the  twist-four operators is nontrivial. 
As well known~\cite{Ferrara:1972xq}, divergence of the leading twist conformal operators $\mathcal{O}_{\{\mu_1\ldots\mu_n\}}$ 
vanishes in a free theory. Using QCD equations of motion (EOM) the operators 
$\partial^{\mu_1}\mathcal{O}_{\{\mu_1\ldots\mu_n\}}$ can be expressed by a sum of contributions of quark-antiquark-gluon operators, e.g. 
\cite{Balitsky:1989ry,Ball:1998ff},
\begin{align}
  \partial^\mu \mathcal O_{\{\mu\alpha\}} &= 2i\bar q g F_{\alpha\mu}\gamma^\mu q\,,
\notag\\
\frac{4}{5} \partial^\mu \mathcal{O}_{\{\mu\alpha\beta\}} & =  -12 i
 \bar q \gamma^\rho \left\{gF_{\rho\beta}\!
\stackrel{\rightarrow}{D}_\alpha
- \stackrel{\leftarrow}{D}_\alpha\! gF_{\rho\beta} +
  (\alpha\leftrightarrow \beta) \right\} q
-4\partial^\rho \bar q (\gamma_\beta g\widetilde{F}_{\alpha\rho} +
\gamma_{\alpha} g\widetilde{F}_{\beta\rho} ) \gamma_5 q
\nonumber\\&\quad 
- \frac{8}{3}\,
\partial_\beta \bar q \gamma^\sigma g\widetilde{F}_{\sigma\alpha} \gamma_5
q - \frac{8}{3}\,\partial_\alpha \bar q \gamma^\sigma
g\widetilde{F}_{\sigma\beta} \gamma_5 q
+\frac{28}{3}\, g_{\alpha\beta} \partial_\rho \bar q \gamma^\sigma
g\widetilde{F}_{\sigma\rho} q\,,
\label{eq:partialOexamples}
\end{align}
etc.
As a consequence, the coefficient functions of these operators in the OPE cannot be found
in the usual way, by considering matrix elements for the on-shell quark states. 
The quark-antiquark-gluon matrix elements can be used instead, but the real problem is to separate the
contribution of interest, given by a particular combination of the quark-antiquark-gluon operators
as in \eqref{eq:partialOexamples},
from the contributions of many other existing quark-antiquark-gluon operators which correspond to 
``dynamical'' power corrections due to quark-gluon correlations. The guiding principle for the separation
of the kinematic and dynamical contributions \cite{Braun:2011zr} is that this separation must be robust
against the factorization scale changes: the contributions that are identified as dynamical cannot be allowed to 
mix with  the kinematic ones, $\langle \partial^{\mu}\mathcal{O}_{\{\mu\mu_1\ldots\mu_n\}}\rangle$ and 
 $\langle \partial^2\mathcal{O}_{\{\mu_1\ldots\mu_n\}}\rangle$, under renormalization.
A systematic approach to isolate kinematic contributions, which does not require explicit
solution of the twist-four renormalization group equations, was developed in~\cite{Braun:2011dg}.
We will use the results of  Ref.~\cite{Braun:2011dg} extensively and refer the reader to this work 
for a derivation and detailed discussion. 

               \subsection{Light-ray operator product expansion}\label{sec:LROPE}

In this section we present the results for the twist expansion of the nonlocal quark-antiquark operator
$\bar q (z_1v)\gamma^\mu q(z_2v)$ retaining kinematic contributions. The Wilson line $[z_1v,z_2v]$ between the quark and 
the antiquark is always implied.
One obtains~\cite{Braun:2011dg} 
\begin{align}
 [ \bar q(z_1v) \gamma^\mu q(z_2v)]_{t2} &= \partial^\mu\int_0^1\!du\, 
[\bar q(z_1uv)\slashed{v}q(z_2uv)]_{lt}\,,
\notag\\
 [ \bar q(z_1v) \gamma^\mu q(z_2v)]_{t3} &=
\frac12 \int_0^1\!udu \!\int_{z_2}^{z_1}\frac{dw}{z_{12}} 
\biggl\{
\Big[(vd)\partial^\mu - (v\partial) d^\mu + v^\mu (d\partial)+ \ln u\, \partial^\mu  v^2 (d \partial) \Big]
\notag\\&\hspace*{3cm}
\times \Big(z_1 [\bar q(z_1uv)\slashed{v}q(wuv)]_{lt} + z_2 [\bar q(wuv)\slashed{v}q(z_2uv)]_{lt}\Big)
\notag\\&\quad
+i\epsilon^{\rho\nu\sigma\mu} v_\rho \partial_\sigma \nabla_\nu  
\Big (z_1 [\bar q(z_1uv)\slashed{v}\gamma_5q(wuv)]_{lt} - z_2 
[\bar q(wuv)\slashed{v}\gamma_5q(z_2uv)]_{lt}\Big)
\biggr\},
\label{t23}
\end{align}
where $z_{12}= z_1-z_2$, $\partial_\mu = \partial/\partial v^\mu$ and $\nabla_\mu$ is the derivative over the total translation,
\begin{align}
 \nabla_\mu \bar q (z_1v)\Gamma q(z_2v) &= \frac{\partial}{\partial y^\mu} \bar q (z_1v +y)\Gamma q(z_2v+y)\big|_{y=0}\,.
\end{align}
For the matrix element $\langle p'|\ldots|p\rangle$ one can replace $\nabla_\mu \mapsto i\Delta_\mu$. 
The leading-twist projection of a nonlocal operator $[\ldots]_{lt}$ is defined as taking into account the 
contribution of the leading-twist operators only in the short-distance expansion. Operationally, this corresponds
to symmetrization and subtraction of all traces and can be implemented in nonlocal form.
\begin{align}
  [ \bar q(z_1v) \slashed{v} q(z_2v)]_{lt} = \Pi(v,n)[ \bar q(z_1n) \slashed{n} q(z_2n)]\,.  
\end{align}
Explicit expressions for the leading twist projection operator $\Pi(v,n)$ in different representations can be found in~\cite{Braun:2011dg}.
Alternatively, one can use that the leading twist projected operator satisfies the d'Alembert equation~\cite{Balitsky:1987bk}
\begin{align}
 \frac{\partial^2}{\partial v^\mu\partial v_\mu}  [ \bar q(z_1v) \slashed{v} q(z_2v)]_{lt} =0\,.
\end{align} 
This equation supplemented with the boundary condition at $v^\mu \to n^\mu$ can be solved using 
a power series ansatz in $v^2$ so that for any function~\cite{Balitsky:1987bk}
\begin{align}
 [f(v)]_{lt} = f(v) - \frac14 v^2 \int_0^1\! \frac{dt}{t}\,  \frac{\partial^2}{\partial v^\mu\partial v_\mu}\, f(tv)  + \mathcal O (v^4)\,.
\label{BB87}
\end{align}
The twist-four contribution is more complicated and can be written as
\begin{align}
  [ \bar q(z_1v) \gamma^\mu q(z_2v)]_{t4} &=
2 v^\mu [A(v;z_1,z_2)]_{lt} + 2 v^2 \partial^\mu [B (v;z_1,z_2)]_{lt}\,,
\label{eq:t4}
\end{align}
where
\begin{align}\label{A1}
A(n;z_1,z_2)&=\frac14\int_0^1du \,\biggl\{ u^2\ln u\,
z_1z_2\, \nabla^2\, \big[ \bar q(uz_1n) \slashed{n} q(uz_2un)\big]
\notag\\
&\quad+\left(z_2\partial_{z_2}-\frac{z_1}{z_{12}}-\ln u\, z_2\partial_{z_2}^2 z_{12}\right) R_2(uz_1n,uz_2n)
\notag\\&\quad
-\left(z_1\partial_{z_1}-\frac{z_2}{z_{21}}-\ln u\, z_1\partial_{z_1}^2 z_{21}\right) \bar R_2(uz_1n,uz_2n)
\biggr\}\,,
\\ 
\label{B1}
B(n;z_1,z_2) &=\frac18\int_0^1\frac{du}{u^2} \,\biggl\{u^2(1\!-\!u^2\!+\!u^2\ln u)\,
z_1z_2\,\nabla^2\, \big[ \bar q(uz_1n) \slashed{n} q(uz_2un)\big]
\notag\\&\quad
 -\left[
(1-u^2)\left(z_2\partial_{z_2}-\frac{z_1}{z_{12}}\right)+(1\!-\!u^2\!+\!u^2\ln u)\,
z_2\partial_{z_2}^2 z_{12}\right] R_2(uz_1n,uz_2n)
\notag\\&\quad
+\left[
(1-u^2)\left(z_1\partial_{z_1}-\frac{z_2}{z_{21}}\right)+(1\!-\!u^2\!+\!u^2\ln u)\,
z_1\partial_{z_1}^2 z_{21}\right] \bar R_2(uz_1n,uz_2n)
\biggl\}
\notag\\&\quad
+ \frac18 \int_0^1\frac{du}{u^2} \,\biggl[  R_1(uz_1n,uz_2n) - \bar  R_1(uz_1n,uz_2n)\biggr].
\end{align}
The $R$-operators are defined as 
\begin{align}\label{RbarR}
R_i(z_1n,z_2n) &= \int_{z_2}^{z_1}dw\, (w-z_2)\, Q_i(z_1n,wn,z_2n)\,,
\notag\\
\bar R_i(z_1n,z_2n)& = \int_{z_2}^{z_1}dw\, (z_1-w)\, \bar Q_i(z_1n,wn,z_2n)\,, \qquad i=1,2\,,
\end{align}
where $Q_1$ and $Q_2$ are the components of twist-four quark-gluon operators with different properties under conformal $SL(2,R)$ transformations
\begin{align}
Q_2(z_1n, w n,z_2n)  - Q_1(z_1n,w n,z_2n)  &=  \bar q(z_1n)  \big[ig F_{+\mu}(wn)  -  g \widetilde F_{+\mu}(wn)\gamma_5\big]\gamma^\mu q(z_2n)\,,
\notag\\
\bar Q_2(z_1n,wn,z_2n)  - \bar Q_1(z_1n,wn,z_2n)  &=  \bar q(z_1n)  \big[ig F_{+\mu}(wn)  + g \widetilde F_{+\mu}(wn)\gamma_5\big]\gamma^\mu q(z_2n)\,.
\end{align}
The operators with and without a ``bar'' are related by hermitian conjugation:
\begin{equation}
 \bar Q_i(z_1n,wn,z_2n)=(Q_i(z_2n,wn,z_1n))^\dagger.
\label{eq:hermiticity}
\end{equation}
Explicit expressions for the $Q_1$, $Q_2$ operators are given in Ref.~\cite{Braun:2011dg} in the two-component spinor representation, Eqs. (2.53) and (2.54).
They will not be needed in what follows.
 
To avoid misunderstanding, we stress that the expressions in \eqref{A1}, \eqref{B1} only include contributions that are relevant for the calculation of kinematic 
power corrections. For this application one can also substitute~\cite{Braun:2011dg}   
\begin{equation}
  Q_1(uz_1n,uwn,uz_2n) \to \frac{w-z_2}{z_{12}}\partial_{z_2} z_{12}Q_2(uz_1n,uwn,uz_2n)
\end{equation} 
and rewrite $R_1$ operators in terms of $Q_2$. 
For our purposes these two representations are equivalent.

Up to overall factors, the expressions in Eqs.~\eqref{A1},\eqref{B1} are very similar to the corresponding
results for the time-ordered product of two electromagnetic currents~\cite{Braun:2011dg}.
The only essential difference is the expression in the last line in \eqref{B1}.
This contribution is equal, up to an overall minus sign, to the contribution 
from gluon emission off the quark propagator connecting 
two currents --- a process absent in our setup. Although this extra term appears simple, it introduces major 
complications into the final result, as demonstrated in the next Section.   
 
\subsection{Short-distance  expansion}\label{sec:SDOPE}

The light-ray quark-antiquark operator (on the light cone) can be expanded in terms of conformal operators~\eqref{ON}:
\begin{align}\label{Oconf-ex}
 \bar q (z_1n)\slashed{n} q(z_2n)&=
\sum_{N=0}^\infty \varkappa_N z_{12}^N\sum_{k=0}^\infty\frac{1}{k!}\int_0^1\!du\, (u\bar u)^{N+1}\,(z_{21}^u)^k
\partial_+^k  \mathcal{O}_{N}(0)\,,
\end{align}
where  $z_{21}^u = \bar u z_2 + u z_1$, $\bar u =1-u$, and
\begin{equation}
 \varkappa_N = \frac{2(2N+3)}{(N+1)!}\,.
\end{equation}
The integral in \eqref{Oconf-ex} in the general case is a hypergeometric function. It simplifies for the choice $z_1 =z$, $z_2 =0$ so that
\begin{align}
 \bar q (zn)\slashed{n} q(0)&=
\sum_{N=0}^\infty 2(2N+3) \sum_{k=0}^\infty \frac{z^{N+k}}{k!}\frac{\Gamma(N+k+2)}{\Gamma(2N+k+4)} \partial_+^k  \mathcal{O}_{N}(0)\,.
\end{align}
These expressions are sufficient to obtain a short-distance expansion of twist-two and twist-three contributions \eqref{t23}.

The twist-four contributions are considerably more complicated. The main observation behind the approach developed
in Ref.~\cite{Braun:2011dg} is that the renormalization group equations (RGEs) for twist-four operators are hermitian 
with respect to a certain scalar product \cite{Braun:2009vc}. The coefficient functions in the expansion of the 
multiplicatively renormalizable twist-four operators in a suitable basis of quark-antiquark-gluon operators 
are, therefore, mutually orthogonal, and one can separate the kinematic operator of interest by an appropriate projection. 
Thus explicit analytic solution of the twist-four RGEs, which is not available, is not needed.  One obtains~\cite{Braun:2011dg}  
\begin{align}
Q_i(z_1n,z_2n,z_3n)&= 
\sum_{N=1}^\infty\sum_{k=0}^\infty C_{Nk}\,
\Psi^{(i)}_{Nk} (z_1,z_2,z_3)\,\partial_+^k(\partial {O})_N  + \ldots
\notag\\
\bar Q_i(z_1n,z_2n,z_3n) &= 
\sum_{N=1}^\infty\sum_{k=0}^\infty C_{Nk}\,
(-1)^{N+1} \Psi^{(i)}_{Nk} (z_3,z_2,z_1)\,\partial_+^k(\partial {O})_N  + \ldots
\label{eq:victory}
\end{align}
where 
\begin{align}
C_{Nk} 
& = 4 \frac{1}{k!} \frac{(2N+3)\Gamma(N+2)}{\Gamma(2 N+4 + k)(N+2)^2\gamma_N}\,, 
\qquad \gamma_N = \psi(N+3)+\psi(N+1)-\psi(3)-\psi(1).
\label{CNk}
\end{align}
Ellipses in \eqref{eq:victory} stand for the contributions of ``dynamical'' twist-four operators. Their matrix elements 
are independent nonperturbative parameters unrelated to the leading-twist GPDs. By construction, the ``dynamical'' contributions 
do not mix with the kinematic ones under renormalization.
In this way the separation between them is well defined: If dynamical contributions are neglected at one particular scale, 
they do not reappear at other scales.
Explicit expressions for the the ``wave functions'' $\Psi^{(i)}_{Nk} (z_1,z_2,z_3)$ and some more details are collected in Appendix~\ref{app:A}. 

Thanks to the identities in \eqref{eq:ident} the expansion for the $R_2$ operator \eqref{RbarR} can be brought to a rather simple form
\begin{align}
R_2(z_1n,z_2n)&=
-\sum_{N=1}^\infty\sum_{k=0}^\infty\frac{\omega_{Nk}}{N+2}
\big(S_+^{(1,0)}\big)^k z_{12}^{N+1}\, \partial_+^k (\partial{O})_N+\ldots\,,
\notag\\
\bar R_2(z_1n,z_2n)&=-\sum_{N=1}^\infty\sum_{k=0}^\infty\frac{\omega_{Nk}}{N+2}
\big(S_+^{(0,1)}\big)^k z_{12}^{N+1}\, \partial_+^k (\partial{O})_N+\ldots\,,
\end{align}
where
\begin{equation}\label{omegaNk}
  \omega_{Nk} = \varkappa_N \frac{1}{k!}\frac{\Gamma(N+2)\Gamma(N+2)}{\Gamma(2N+4+k)}\,
\end{equation}
and
\begin{eqnarray}
  (S_+^{(j_1,j_2)})^k z_{12}^N = z_{12}^N \frac{\Gamma[2N+2j_1+2j_2+k]}{\Gamma[N+2j_1]\Gamma[N+2j_2]}
  \int_0^1\!dt\, t^{N+2j_1-1}\bar t^{N+2j_2-1}\, (z_{21}^t)^k.
\end{eqnarray}
The differential operators $S_+^{j_1,j_2}$ are defined in Eq.~\eqref{Splus}.
Crucially, the factor  $1/\gamma_N$ in Eq.~\eqref{CNk} has disappeared.
 
The short-distance expansion for the $R_1$ operator \eqref{RbarR} is much more involved. 
We did this calculation for the particular choice $z_1=z$ and $z_2=0$ which is favored in lattice
calculations. For this case we obtain
\begin{align}
 R_1(z n, 0) &
= \sum_{N=1}^\infty\sum_{k=0}^\infty C_{Nk} S_{Nk} z^{N+k+1} \partial_+^k (\partial{O})_N+\ldots 
\notag\\
 \bar R_1(z n, 0) &
= \sum_{N=1}^\infty\sum_{k=0}^\infty C_{Nk} \bar S_{Nk} z^{N+k+1} \partial_+^k (\partial{O})_N+\ldots 
\end{align}
where the coefficients $C_{Nk}$ are defined in \eqref{CNk}, and 
\begin{align}
 S(N,k) &= 
\frac12
(N+2)\sum_{m=0}^{N-1}
\biggl\{
\frac{(-1)^{N-m+1} \Gamma(N+2)}{\Gamma(m+2)\Gamma(N-m+2)}
+ 1
- \frac{(N+3)(N-m)}{(N+1)(N+2)}
\biggr\}
\notag\\&\quad\times
 \frac{\Gamma(m+k+2)}{\Gamma(m+1)} \frac{1}{N-m+1}
 {}_3F_2\left(\begin{matrix}-k,N-m+1,N-m+1 \\ -m-k-1,N-m+2 \end{matrix}; 1 \right),
\notag\\
\widebar{S}(N,k) &=
 (N+2)\sum_{m=0}^{N-1} \biggl\{ 
\frac{(-1)^{N-m+1}\Gamma(N+1)}{\Gamma(m+1) \Gamma(N-m+3)}
+ \frac12
\biggr\} \biggl\{
\frac{(m+1)}{(N-m)} \frac{\Gamma(N+k+2)}{\Gamma(N+2)} 
\notag\\&
-  \frac{\Gamma(m+k+2)}{\Gamma(m+1)}  \frac{1}{N-m+1}
 {}_3F_2\left(
\begin{matrix}
-k,N-m+1,N-m+1\\
-m-k-1,N-m+2
\end{matrix};1
\right)
\biggr\}.
\label{SbarS}
\end{align}
The remaining calculation is simple as the $u$-integrations in \eqref{A1}, \eqref{B1} only produce overall factors.
Note that the operator  $\nabla^2\, \big[ \bar q(uz_1n) \slashed{n} q(uz_2un)\big]$ in the first line in these equations does not 
contribute for $z_2=0$ or  $z_1=0$. 

               \subsection{Matrix elements}\label{sec:ME}

The GPDs \eqref{defNucleonGPD} are defined as matrix elements of the light-ray operators, and enter the OPE through the leading
twist projection of the quark-antiquark operators at non-light-like separations. Note also that the short-distance expansion 
of nonlocal operators involves local operators with additional total derivatives \eqref{Oconf-ex}, $\partial_+ \mapsto i\Delta_+$, so that we need a slightly
more general expression as compared to \eqref{iO}%
\footnote{The notation in \eqref{iOv} is somewhat misleading as the prefactors $\bar u(p')\slashed{v}u(p)$ for $M$ and $(Pv)$ for $E$ have to be included in the leading twist
projection.}: 
\begin{align}\label{iOv}
(-i)^N\langle p'|[(\Delta v)^k {O}^{(\slashed{v})}_N]_{lt} |p\rangle&=
\bar u(p')\slashed{v}u(p)\, [(\Delta v)^kM_{N}^{(v)}(t)]_{lt} - \frac{Pv}{m_N}\bar u(p')u(p)\,[(\Delta v)^kE_{N}^{(v)}(t)]_{lt}\,.
\end{align}
Using the explicit expression for the projection operator $\Pi(v,n)$~\cite{Braun:2011dg}, or, alternatively, using  Eq.~\eqref{BB87}, one 
obtains to $\mathcal{O}(v^2)$ accuracy
\begin{align}\label{888V}
{}[(\Delta v)^k M^{(v)}_N]_{lt}&=
\sum_{\substack{m=0,\\ \text{even}}}^N (\Delta v)^{m+k} (P v)^{N-m} M_{Nm}
-\frac{v^2}{4(N\!+\!k\!+\!1)}\!\sum_{\substack{m=-2,\\ \text{even}}}^{N-2}
(\Delta v)^{m+k}(Pv)^{N-m-2} \mathbb M^{(4)}_{Nkm},
\notag\\
{}[(\Delta v)^kE_N^{(v)}]_{lt}&=\sum_{\substack{m=0,\\ \text{even}}}^{N+1}(\Delta v)^{m+k} (P v)^{N-m}E_{Nm} 
-\frac{v^2}{4(N\!+\!k\!+\!1)}
\sum_{\substack{m=-2,\\ \text{even}}}^{N-1}
(\Delta v)^{m+k}(Pv)^{N-m-2}\mathbb E^{(4)}_{Nkm},
\end{align}
where
\begin{align}
\mathbb M^{(4)}_{Nkm} &= (m\!+\!k\!+\!1)(m\!+\!k\!+\!2)
t\, M_{Nm+2}+(N\!-\!m)(N\!-\!m\!-\!1)\Big(m_N^2-\frac{t}4\Big)M_{Nm}\,,
\notag\\
\mathbb E^{(4)}_{Nkm} &= (m\!+\!k\!+\!1)(m\!+\!k\!+\!2)
t\,E_{Nm+2}+ (N\!-\!m)(N\!-\!m\!+\!1)\Big(m_N^2-\frac{t}4\Big)E_{Nm}
\notag\\&\quad
-  2(N\!-\!m)m_N^2 M_{Nm}\,.
\label{Mnkm-Enkm}
\end{align}
The sum in \eqref{888V}  goes over even values of $m$, and it is tacitly assumed that the generalized form factors $M_{Nm}$, $E_{Nm}$ vanish for $m<0$. 
Here and below the $t$ dependence of  $M_{Nm}$, $E_{Nm}$ is not shown for brevity.

The terms $\mathcal{O}(v^2)$ in \eqref{iOv} give rise to finite-$t$ and target mass corrections to quasi-GPDs, which are analogous to the 
Nachtmann corrections to deep inelastic scattering. Main contribution of this work is the calculation of corrections due to the ``kinematic'' twist-four
operator $(\partial O)_N$ which must be taken into account alongside the Nachtmann corrections to maintain Lorentz invariance to twist-four accuracy. 
The corresponding matrix elements can be calculated using that 
\begin{align}
 \langle p'|(\partial O^{(\slashed{v})})_N  |p\rangle&= \frac{1}{N+1}i\Delta^\mu \frac{\partial}{\partial v^\mu} 
\langle p'|[ {O}^{(\slashed{v})}_N]_{lt} |p\rangle\,.
\end{align}
Note that one has to use the leading twist projection of the off-light-cone matrix element on the r.h.s. 
since taking the derivative $\partial/\partial n^\mu$ is in conflict with the constraint $n^2=0$.
One obtains 
\begin{align}\label{partialOME}
(-i)^{N-1}\langle p'|(\partial {O})^{(\slashed{v})}_N |p\rangle&=
\bar u(p')\slashed{v}u(p)\, (\partial M)_{N}  - \frac{Pv}{m_N}\bar u(p')u(p)\,(\partial E)_N\,,
\notag\\
(\partial M)_N &= \frac{1}{2(N\!+\!1)^2} \sum_{\substack{m=0,\\ \text{even}}}^{N-2}(\Delta v)^{m+1} (P v)^{N-m-2} (\partial\mathbb M)_{Nm}\,,
\notag\\
 (\partial E)_N &= \frac{1 }{2 (N+1)^2} \sum_{\substack{m=0,\\ \text{even}}}^{N-1} (\Delta v)^{m+1} (P v)^{N-m-2} (\partial\mathbb E)_{Nm}\,,  
\end{align}
where
\begin{align}
(\partial\mathbb M)_{Nm} &=
(m\!+\!2)(2N\!-\!m\!+\!1) t M_{Nm+2}
- (N\!-\!m)(N\!-\!m\!-\!1)\left(m_N^2-\frac{t}4\right)M_{Nm}\,,
\notag\\
(\partial\mathbb E)_{Nm} &=
(m+2)(2N-m+1)\, t E_{Nm+2}
- (N-m)(N-m+1)\left(m_N^2-\frac{t}4\right)E_{Nm}
\notag\\&\quad
{+} 2(N-m)m_N^2 M_{Nm}\,.
\end{align}
Since these operators are already twist-four, one can replace in the matrix elements $n_\mu\leftrightarrow v_\mu$ without additional kinematic corrections.

The expressions for the matrix elements of axial-vector operators are very similar. For the leading-twist operator one obtains
\begin{align}\label{iOv}
(-i)^N\!\langle p'|[(\Delta v)^k {O}^{(\slashed{v}\gamma_5)}_N]_{lt} |p\rangle&=
\bar u(p')\slashed{v}\gamma_5u(p)\, [(\Delta v)^k\widetilde H_{N}^{(v)}(t)]_{lt} \!+\! \frac{Pv}{ 2 m_N}\bar u(p')\gamma_5u(p)\,[(\Delta v)^k\widetilde E_{N}^{(v)}(t)]_{lt}
\end{align}
with
\begin{align}\label{888A}
{}[(\Delta v)^k \widetilde H^{(v)}_N]_{lt}&=
\sum_{\substack{m=0,\\ \text{even}}}^N (\Delta v)^{m+k} (P v)^{N-m} \widetilde H_{Nm}
-\frac{v^2}{4(N\!+\!k\!+\!1)}\sum_{\substack{m=-2,\\ \text{even}}}^{N-2}
(\Delta v)^{m+k}(Pv)^{N-m-2} \widetilde{\mathbb H}^{(4)}_{Nm},
\notag\\
{}[(\Delta v)^k\widetilde E_N^{(v)}]_{lt}&=\sum_{\substack{m=0,\\ \text{even}}}^{N+1}(\Delta v)^{m+k} (P v)^{N-m}\widetilde E_{Nkm} 
-\frac{v^2}{4(N\!+\!k\!+\!1)}\!\sum_{\substack{m=-2,\\ \text{even}}}^{N-1}
(\Delta v)^{m+k}(Pv)^{N-m-2}\widetilde{\mathbb E}^{(4)}_{Nkm},
\end{align}
where
\begin{align}
\widetilde{\mathbb H}^{(4)}_{Nkm} &= (m\!+\!k\!+\!1)(m\!+\!k\!+\!2)
t\, \widetilde H_{Nm+2}+(N\!-\!m)(N\!-\!m\!-\!1)\Big(m_N^2-\frac{t}4\Big)\widetilde H_{Nm}\,,
\notag\\
\widetilde{\mathbb E}^{(4)}_{Nkm} &= (m\!+\!k\!+\!2)(m\!+\!k\!+\!3)
t\,\widetilde E_{Nm+2}+ (N\!-\!m)(N\!-\!m\!-\!1)\Big(m_N^2-\frac{t}4\Big)\widetilde E_{Nm}
\notag\\&\quad
+ 8 (m\!+\!k\!+\!2)m_N^2 \widetilde H_{Nm+2}\,.
\label{tildeMathbbHE}
\end{align}
The matrix elements of the axial-vector ``kinematical'' twist-four operator are equal to
\begin{align}\label{partialOMEAxial}
(-i)^{N-1}\langle p'|(\partial {O}^{(\slashed{v}\gamma_5)})_N |p\rangle&=
\bar u(p')\slashed{v}\gamma_5u(p)\, (\partial \widetilde H)_{N}  + \frac{(\Delta v)}{2 m_N}\bar u(p')\gamma_5u(p)\,(\partial \widetilde E)_N\,,
\notag\\
  (\partial \widetilde H)_{N} &= 
 \frac{1}{2(N+1)^2}\sum_{\substack{m=0,\\ \text{even}}}^{N-2} (\Delta v)^{m+1}(P v)^{N-m-2} (\partial\widetilde{\mathbb H})_{Nm}\,,
\notag\\
  (\partial \widetilde E)_{N} &= 
 \frac{1}{2(N+1)^2} \sum_{\substack{m=0,\\ \text{even}}}^{N-2}  (\Delta x)^{m+1} (Px)^{N-m-2} (\partial\widetilde{\mathbb H})_{Nm}\,,
\end{align}
where
\begin{align}
(\partial\widetilde{\mathbb H})_{Nm} &= 
 (m+2)(2N-m+1) t \widetilde{H}_{Nm+2} - (N\!-\!m)(N\!-\!m\!-\!1)\Big(m_N^2-\frac{t}4\Big)\widetilde H_{Nm}\,,
\notag\\
(\partial\widetilde{\mathbb E})_{Nm}&=
(m+3)(2N-m)  t\,\widetilde E_{Nm+2} -  (N\!-\!m)(N\!-\!m\!-\!1)\Big(m_N^2-\frac{t}4\Big)\widetilde E_{Nm}
\notag\\&\quad
+ 8(N-m-1)m_N^2 \widetilde H_{Nm+2}\,.
\label{partialOMEAxial2}
\end{align}


               \section{Results}\label{sec:results}


Combining  the light-ray OPE results in Sec.~\ref{sec:LROPE}, the short-distance expansion in Sec.~\ref{sec:SDOPE}
and the matrix elements in Sec.~\ref{sec:ME}, the final expressions for the finite-$t$ and target mass corrections
can be obtained by simple algebra. In this section we  present the results for the short-distance expansion of the invariant amplitudes
in terms of  Gegenbauer moments of the GPDs \eqref{iO}, \eqref{moexp}.  
We use the Lorentz/Dirac structure decomposition defined in Sec.~\ref{sec:defqGPD}, Eqs.~\eqref{Vstructures}, \eqref{Astructures},
separating the contributions of the operators of different twist 
\begin{align}
 \langle p'|\bar\psi(z v) \gamma^\mu \psi(0)|p\rangle &= 
 \langle p'|[ \bar\psi(z v) \gamma^\mu \psi(0)]_{t2}|p\rangle 
 + \langle p'|[ \bar\psi(z v) \gamma^\mu \psi(0)]_{t3}|p\rangle 
\notag\\&\quad
 + \langle p'|[ \bar\psi(z v) \gamma^\mu \psi(0)]_{t4}|p\rangle + \ldots 
\end{align}
For convenience, the results for the first few terms in the short-distance expansion, $\sim z^N$, $N=0,1,2,3$, summed over twists, are collected in the Appendix.
We use the notations
\begin{align}
 P_v = (Pv)\,,\qquad \Delta_v = (\Delta v)\,, \qquad \eta = - \frac{\Delta_v}{2P_v}\,.
\end{align} 
We refer to the $\eta$-variable as the  lattice asymmetry parameter~\eqref{eta}.

               \subsection{Twist-two}\label{sec:t2}

The finite-$t$ and target mass corrections originating from the matrix elements of the leading twist-two operators
present a straightforward generalization of Nachtmann target mass corrections to the deep-inelastic scattering.
We write the results as
\begin{align}
 (\mathcal F)_{t2} &= 
 \sum_{N=0}^\infty 2(2N+3) \sum_{k=0}^\infty \frac{(iz P_v)^{N+k}}{k!}
\frac{\Gamma(N+k+1)}{\Gamma(2N+k+4)}  
(\mathcal F_{Nk})_{t2}\,,
\end{align}
where \eqref{Vstructures}  
\begin{align}
 \mathcal F \in \{  \mathcal M^{(v)}, \mathcal M^{(P)}, \mathcal M^{(\Delta)}, \mathcal M^{(\gamma)}, \mathcal E^{(v)}, \mathcal E^{(P)}, \mathcal E^{(\Delta)}\}.
\label{TwistTwoSet}
\end{align}
One obtains
\begin{align}
 (\mathcal{M}^{(v)}_{Nk})_{t2} &=
 (N+k+1) \sum_{\substack{m=0,\\ \text{even}}}^N (-2\eta)^{m+k} M_{Nm}
-\frac14 \frac{v^2}{P_v^2}
\sum_{\substack{m=-2,\\ \text{even}}}^{N-2}
(-2\eta)^{m+k}\,\mathbb M^{(4)}_{Nkm}, 
\notag\\
 (\mathcal{M}^{(P)}_{Nk})_{t2} &=
(N+k+1) \sum_{\substack{m=0,\\ \text{even}}}^N(-2\eta)^{m+k} M_{Nm}
- \frac{(N+k-1)}{4(N+k+1)} \frac{v^2}{P_v^2} \sum_{\substack{m=-2,\\ \text{even}}}^{N-2}
(-2\eta)^{m+k}\, \mathbb M^{(4)}_{Nkm}, 
\notag\\
 (\mathcal{M}^{(\Delta)}_{Nk})_{t2} &=
\sum_{\substack{m=0,\\ \text{even}}}^N (m\!+\!k) (-2\eta)^{m+k-1} M_{Nm}
-\frac{1}{4(N\!+\!k\!+\!1)} \frac{v^2}{P_v^2}  \sum_{\substack{m=-2,\\ \text{even}}}^{N-2}
(m\!+\!k) (-2\eta)^{m+k-1}\, \mathbb M^{(4)}_{Nkm}, 
\notag\\
(\mathcal{M}^{(\gamma)}_{Nk})_{t2} &=
\sum_{\substack{m=0,\\ \text{even}}}^N(-2\eta)^{m+k} M_{Nm}
-\frac{1}{4(N+k+1)}  \frac{v^2}{P_v^2} 
\sum_{\substack{m=-2,\\ \text{even}}}^{N-2}
(-2\eta)^{m+k}\, \mathbb M^{(4)}_{Nkm}\,,
\end{align}
and
\begin{align}
 (\mathcal{E}^{(v)}_{Nk})_{t2} &=
 (N+k+1) \sum_{\substack{m=0,\\ \text{even}}}^{N+1} (-2\eta)^{m+k}  E_{Nm} 
 -\frac{1}{4} \frac{v^2}{P_v^2}\sum_{\substack{m=-2,\\ \text{even}}}^{N-1} (-2\eta)^{m+k}\, \mathbb E^{(4)}_{Nkm},
\nonumber\\
 (\mathcal{E}^{(P)}_{Nk})_{t2} &=
 (N+k+1) \sum_{\substack{m=0,\\ \text{even}}}^{N+1} (-2\eta)^{m+k} E_{Nm}
- \frac{(N+k-1)}{4(N+k+1)}\frac{v^2}{P_v^2} 
\sum_{\substack{m=-2,\\ \text{even}}}^{N-1} (-2\eta)^{m+k} \, \mathbb E^{(4)}_{Nkm}, 
\notag\\
 (\mathcal{E}^{(\Delta)}_{Nk})_{t2} &=
\sum_{\substack{m=0,\\ \text{even}}}^{N+1}
  (m\!+\!k) (-2\eta)^{m+k-1} E_{Nm} 
-\frac{1}{4(N\!+\!k\!+\!1)}\frac{v^2}{P_v^2}
 \sum_{\substack{m=-2,\\ \text{even}}}^{N-1}
(m\!+\!k) (-2\eta)^{m+k-1}  \mathbb E^{(4)}_{Nkm}, 
\end{align}
where $ \mathbb M^{(4)}_{Nkm} $ and  $\mathbb E^{(4)}_{Nkm}$ are defined in Eq.~\eqref{Mnkm-Enkm}. 

               \subsection{Twist-three}\label{sec:t3}

Twist-three contributions can be written as
\begin{align}
 (\mathcal F)_{t3} &= 
 \sum_{N=0}^\infty  \frac{(2N+3)}{(N+1)(N+2)} \sum_{k=0}^\infty \frac{(iz P_v)^{N+k+1}}{k!}
 \frac{\Gamma(N+k+2)}{\Gamma(2N+k+4)}
(\mathcal F_{Nk})_{t3}
\end{align}
with the invariant amplitudes from the set 
\begin{align}
 \mathcal F \in \{  \mathcal M^{(v)}, \mathcal M^{(P)}, \mathcal M^{(\Delta)}, \mathcal M^{(\gamma)}, \mathcal E^{(v)}, \mathcal E^{(P)}, \mathcal E^{(\Delta)},
 \widetilde{\mathcal H}^{(\gamma)}, \widetilde{\mathcal H}^{(\tilde \Delta)},  \widetilde{\mathcal E}^{(\tilde \Delta)}\}.
\label{TwistThreeSet}
\end{align}
The first seven amplitudes \eqref{Vstructures} contain contributions of the vector GPDs $M$ and $E$, and the last three ones involve the 
axial-vector GPDs $\widetilde H$ and $\widetilde E$. 
Note that the twist-three matrix elements satisfy the constraints
\begin{align}
v_\mu \langle p'|[ \bar\psi(z v) \gamma^\mu \psi(0)]_{t3}|p\rangle = 0\,, \qquad \frac{\partial}{\partial v^\mu}\langle p'|[ \bar\psi(z v) \gamma^\mu \psi(0)]_{t3}|p\rangle =0\,. 
\end{align}
As a consequence,  $(\mathcal M^{(v)})_{t3}$  and  $(\mathcal E^{(v)})_{t3}$ amplitudes vanish identically. For the remaining ones we obtain
\allowdisplaybreaks
\begin{align}
 (\mathcal{M}^{(P)}_{Nk})_{t3} &=
\frac{1}{(N+k+2)}\frac{v^2}{P_v^2}\sum_{\substack{m=-2,\\ \text{even}}}^{N-2} (-2\eta)^{m+k+1}  
\biggl\{\mathbb M^{(4)}_{Nkm} - (N+k) (m+k+2) t M_{Nm+2}\biggr\},
\nonumber\\
 (\mathcal{M}^{(\Delta)}_{Nk})_{t3} &=
- \!\!\sum_{\substack{m=0,\\ \text{even}}}^N (N\!-\!m\!+\!1) (-2\eta)^{m+k} M_{Nm}
 + \frac{1}{4(N\!+\!k\!+\!2)}\frac{v^2}{P_v^2}\sum_{\substack{m=-2,\\ \text{even}}}^{N-2} (N\!-\!m\!-\!1)(-2\eta)^{m+k}
\notag\\&\qquad\qquad
\times \biggl\{\frac{N+k}{N+k+1} \mathbb M^{(4)}_{Nkm}
+4 (N-m) \Big(m_N^2-\frac{t}4\Big)  M_{Nm}\biggr\}\,,
\nonumber\\
 (\mathcal{M}^{(\gamma)}_{Nk})_{t3} &=
\sum_{\substack{m=0,\\ \text{even}}}^N (-2\eta)^{m+k+1} M_{Nm}
-\frac{1}{4(N\!+\!k\!+\!2)}\frac{v^2}{P_v^2}\sum_{\substack{m=-2,\\ \text{even}}}^{N-2}
(-2\eta)^{m+k+1}
\notag\\&\qquad\qquad
\times\biggl\{
\frac{(N+k)}{(N\!+\!k\!+\!1)}\mathbb M^{(4)}_{Nkm}
+ 4 (m\!+\!k\!+\!2) t M_{Nm+2}\biggr\}, 
\end{align}
\begin{align}
 (\mathcal{E}^{(P)}_{Nk})_{t3} &=
\frac{1}{(N+k+2)} \frac{v^2}{P_v^2}\sum_{\substack{m=-2,\\ \text{even}}}^{N-1} (-2\eta)^{m+k+1} 
\biggl\{\mathbb E^{(4)}_{Nkm} - (N+k) (m+k+2)  t E_{Nm+2}\biggr\},
\nonumber\\
 (\mathcal{E}^{(\Delta)}_{Nk})_{t3} &=
%
%
- \sum_{\substack{m=0,\\ \text{even}}}^{N+1} (N\!-\!m\!+\!1) (-2\eta)^{m+k} E_{Nm} 
\notag\\&\quad
+\frac{1}{4(N\!+\!k\!+\!2)} \frac{(N+k)}{(N\!+\!k\!+\!1)} \frac{v^2}{P_v^2} \sum_{\substack{m=-2,\\ \text{even}}}^{N-1}
(N\!-\!m\!-\!1) (-2\eta)^{m+k} \mathbb E^{(4)}_{Nkm} 
\notag\\&\quad
+\frac{1}{(N\!+\!k\!+\!2)} \frac{v^2}{P_v^2} \sum_{\substack{m=-2,\\ \text{even}}}^{N-2}
 (N\!-\!m)(-2\eta)^{m+k} \Big[
(N\!-\!m\!+\!1)\Big(m_N^2-\frac{t}4\Big)E_{Nm}
- 2 m_N^2 M_{Nm}
\Big]
\end{align}
and finally
\begin{align}
(\widetilde{\mathcal H}^{(\gamma)}_{Nk})_{t3} &= \phantom{-}
 2 \sum_{\substack{m=0,\\ \text{even}}}^N(-2\eta)^{m+k} \widetilde{H}_{Nm}
 -  \frac{1}{2(N+k+1)} \frac{v^2}{P_v^2}  \sum_{\substack{m=-2,\\ \text{even}}}^{N-2}  (-2\eta)^{m+k}\,\widetilde{\mathbb{H}}^{(4)}_{Nkm}\,, 
\notag\\
(\widetilde{\mathcal H}^{(\tilde \Delta)}_{Nk})_{t3} &=
- \sum_{\substack{m=0,\\ \text{even}}}^N(N\!-\!m) (-2\eta)^{m+k} \widetilde{H}_{Nm}
+  \frac{1}{4}  \frac{v^2}{P_v^2}\sum_{\substack{m=-2,\\ \text{even}}}^{N-2}\frac{(N\!-\!m\!-\!2)}{(N\!+\!k\!+\!1)}  (-2\eta)^{m+k}\,\widetilde{\mathbb{H}}^{(4)}_{Nkm}\,, 
\notag\\
(\widetilde{\mathcal E}^{(\tilde \Delta)}_{Nk})_{t3} &=
- \sum_{\substack{m=0,\\ \text{even}}}^N(N\!-\!m) (-2\eta)^{m+k} \widetilde{E}_{Nm}
+  \frac{1}{4}  \frac{v^2}{P_v^2} \sum_{\substack{m=-2,\\ \text{even}}}^{N-2}\frac{(N\!-\!m\!-\!2)}{(N\!+\!k\!+\!1)}  (-2\eta)^{m+k}\,\widetilde{\mathbb{E}}^{(4)}_{Nkm} \,,
\end{align}
where $\widetilde{\mathbb{H}}^{(4)}_{Nkm}$ and $\widetilde{\mathbb{E}}^{(4)}_{Nkm}$ are defined in Eq.~\eqref{tildeMathbbHE}.

               \subsection{Twist-four}\label{sec:t4}

Twist-four contributions are more involved:
\begin{align}
 (\mathcal F)_{t4} &= 
 \frac12 \sum_{N=0}^\infty  \frac{(2N+3)}{(N+1)^2(N+2)^2} \sum_{k=0}^\infty \frac{(iz P_v)^{N+k+1}}{k!}
 \frac{\Gamma(N+k+2)}{\Gamma(2N+k+4)}
\biggl[
 (\mathcal F^A_{Nk})_{t4}
+ T_{Nk} (\mathcal F^B_{Nk})_{t4}\biggr],
\end{align}
where $(\mathcal F^A_{Nk})_{t4}$ and  $(\mathcal F^B_{Nk})_{t4}$ correspond to the contributions in 
Eqs.~\eqref{A1} and \eqref{B1}, respectively, and 
\begin{align}
T_{Nk} &= \frac{1}{N+k}\biggl\{\frac12 N(N+3) 
+ \frac{\Gamma(N+2)}{\Gamma(N+k+2)} \frac{S(N,k)- \bar S(N,k)}{\gamma_N} \biggr\}.
\label{TNk}
\end{align}
We obtain
\begin{align}
 (\mathcal{M}^{(v),A}_{Nk})_{t4} &=
  \frac{v^2}{P_v^2} \sum_{\substack{m=0,\\ \text{even}}}^{N-2}(-2\eta)^{m+k+1} (\partial\mathbb M)_{Nm}\,,
\notag\\
 (\mathcal{M}^{(v),B}_{Nk})_{t4} & =  (\mathcal{M}^{(P),B}_{Nk})_{t4} = 
 \frac{v^2}{P_v^2} \sum_{\substack{m=0,\\ \text{even}}}^{N-2} (N+k) (-2\eta)^{m+k+1}  (\partial\mathbb M)_{Nm}\,,
\notag\\
 (\mathcal{M}^{(\Delta),B}_{Nk})_{t4} & =
 \frac{v^2}{P_v^2} \sum_{\substack{m=0,\\ \text{even}}}^{N-2} (m+k+1) (-2\eta)^{m+k}  (\partial\mathbb M)_{Nm}\,,
\notag\\
 (\mathcal{M}^{(\gamma),B}_{Nk})_{t4} & =
  \frac{v^2}{P_v^2} \sum_{\substack{m=0,\\ \text{even}}}^{N-2} (-2\eta)^{m+k+1}  (\partial\mathbb M)_{Nm}\,,
\end{align}
and
\begin{align}
 (\mathcal{E}^{(v),A}_{Nk})_{t4} & =
 \frac{v^2}{P_v^2} \sum_{\substack{m=0,\\ \text{even}}}^{N-1} (-2\eta)^{m+k+1} (\partial\mathbb E)_{Nm}\,,  
\notag\\
 (\mathcal{E}^{(v),B}_{Nk})_{t4} & = (\mathcal{E}^{(P),B}_{Nk})_{t4} =
 \frac{v^2}{P_v^2} \sum_{\substack{m=0,\\ \text{even}}}^{N-1}  (N+k) (-2\eta)^{m+k+1}  (\partial\mathbb E)_{Nm}\,,
\notag\\
(\mathcal{E}^{(\Delta),B}_{Nk})_{t4} &=
 \frac{v^2}{P_v^2} \sum_{\substack{m=0,\\ \text{even}}}^{N-1} (m+k+1) (-2\eta)^{m+k}  (\partial\mathbb E)_{Nm}\,.
\end{align}
Note that the $A$-structure \eqref{A1} only contributes to $(\mathcal{M}^{(v),A}_{Nk})_{t4}$ and $(\mathcal{E}^{(v),A}_{Nk})_{t4}$.


               \section{$\mathcal{M}^{(P)}$ amplitude: the first few moments and numerical estimates}\label{sec:MP}


The invariant amplitude $\mathcal{M}^{(P)}$  \eqref{Vstructures} is likely to be 
the most suitable one for lattice applications. In this section, we present the results
for the first few terms in the expansion of this amplitude in powers of the ``Ioffe time'' 
$z(P\cdot v)$ using the normalization specified in Eq.~\eqref{SDE}.
Furthermore, we present estimates for the size of finite-$t$ and nucleon mass corrections 
for this amplitude using a simple GPD model. 
The expressions for the first four moments for the all invariant amplitudes are 
collected in appendix~\ref{app:B}. 

We remind that the ``magnetic'' GPD $M$ is defined as $M=H+E$. In what follows we do not include the 
contributions of axial-vector GPDs $\widetilde H$ and $\widetilde E$ which are collected as 
separate amplitudes \eqref{Astructures}. 

The first two terms in the expansion, $N=0,1$, correspond to the operators with zero and one covariant
derivatives. They do not receive any kinematic power corrections,
\begin{align}
 \mathcal{M}^{(P)}_0 &= M_{00}(t)\,,
\notag\\
 \mathcal{M}^{(P)}_1 &= M_{10}(t) - 6\eta M_{00}(t)\,.
\end{align}
The $N=2$ term  is already affected. There are Nachtmann-type power corrections 
from the contribution of the twist-two operators and the correction due to the twist-three 
operators
\begin{align}
 (\mathcal{M}^{(P)}_2)_{t2} &= M_{20}(t) -10\eta M_{10}(t)  + \eta^236 \eta^2 M_{00}(t)+4 \eta^2 M_{22}(t)
\notag\\&\quad
- \frac{1}{18}\frac{v^2}{P_v^2} \Big[P^2 M_{20}(t)+  9 t M_{00}(t) + t M_{22}(t)\Big]\,,  
\notag\\
 (\mathcal{M}^{(P)}_2)_{t3} &= - \frac52 \frac{v^2}{P_v^2} t M_{00}\,, 
\label{M20t2t3}
\end{align} 
so that in the sum
\begin{align}
\mathcal{M}^{(P)}_2 &= 
M_{20}(t) -10\eta M_{10}(t)  + 36 \eta^2 M_{00}(t)+4 \eta^2 M_{22}(t)
\notag\\&\quad
- \frac{1}{18}\frac{v^2P^2}{P_v^2} M_{20}(t) 
- \frac{1}{18}\frac{v^2t }{P_v^2} \Big[54 M_{00}(t) + M_{22}(t)\Big]\,.  
\end{align}
Here and below $P^2 = m_N^2-t/4$, cf.~\eqref{P2}. 
Note that the $\mathcal{O}(t)$ corrections are much larger than the  $\mathcal{O}(P^2)$ ones,
but are somewhat moderated by the expected different $t$-dependence of the form factors 
$M_{00}(t)$ and $M_{20}(t)$, as we will see below. 
  
The ``kinematic'' twist-four operators contribute starting $N=3$%
\footnote{For the $\mathcal{E}^{(P)}$ amplitude, the twist-four contributions appear for $N=2$ already,
same as for a scalar target, see Sec.~\ref{sec:example}.}.
We obtain
\begin{align}
 (\mathcal{M}^{(P)}_3)_{t2} &= M_{30}(t) - 14\eta M_{20}(t) +  4\eta^2\big(20 M_{10}(t)\!+\!M_{32}(t)\big)
 -56\eta^3\big(4M_{00}(t)\!+\!M_{22}(t)\big)
\notag\\&\quad
-\frac{1}{16} \frac{v^2P^2}{P_v^2}
\Big[3M_{30}(t) -  14\eta M_{20}(t)\Big]
\notag\\&\quad
-\frac{1}{16} \frac{v^2 t}{P_v^2}
\Big[20M_{10}(t)+  M_{32}(t) - 42\eta\big(4M_{00}(t)+M_{22}(t)\big)\Big],
\notag\\
 (\mathcal{M}^{(P)}_3)_{t3} &=
-\frac{7}{12} \frac{v^2P^2}{P_v^2} \eta M_{20}(t)
-\frac{7}{12} \frac{v^2t }{P_v^2}
\Big[ 5 M_{10}(t) - \eta  \big(54 M_{00}(t)+M_{22}(t)\big)\Big],
\notag\\
 (\mathcal{M}^{(P)}_3)_{t4} &=
\frac{35}{72}\frac{v^2P^2}{P_v^2} \eta M_{20}(t) 
-  \frac{175}{72}\frac{v^2t}{P_v^2}\eta M_{22}(t)\,, 
\label{M30t2t3t4}
\end{align}
and in the sum of all twists
\begin{align}
\mathcal{M}^{(P)}_3 &= M_{30}(t) - 14\eta M_{20}(t) +  4\eta^2\big(20 M_{10}(t)\!+\!M_{32}(t)\big)
 -56\eta^3\big(4M_{00}(t)\!+\!M_{22}(t)\big)
\notag\\&\quad
-  \frac{1}{144} \frac{v^2P^2}{P_v^2} 
\Big[
 27 M_{30}(t)  -  112 \eta M_{20}(t)
\Big]
\notag\\&\quad
- \frac{1}{144} \frac{v^2 t}{P_v^2}
\Big[
 600 M_{10}(t)+9 M_{32}(t) - 112\eta\big(54  M_{00}(t) + M_{22}(t) \big)
\Big]. 
\end{align}
Also in this case the coefficients of  $\mathcal{O}(t)$ corrections 
are much larger than of the  $\mathcal{O}(P^2)$ ones.
Note also that the contribution of twist-four operator is proportional to $\eta$ and vanishes 
in the zero lattice asymmetry limit $\eta=0$. This feature is general and holds for all moments.

For numerical estimates, we use a simple valence quark GPD model 
based on the standard double-distribution (DD) ansatz~\cite{Radyushkin:1997ki} 
with the $t$ dependence suggested by the Regge calculus~\cite{Kroll:2012sm} 
\begin{align}
 M^q(x,\xi,t) &= \iint\limits_{|\beta|+|\alpha|\le 1}d\alpha d\beta\,
\delta(x-\beta -\xi \alpha) f^q(\beta,\alpha,t)\,,
\end{align}
where 
\begin{align}
 f^q(\beta,\alpha,t) &= q(\beta,t) h(\beta,\alpha)
\notag\\
q(\beta,t) &= \theta(\beta) \beta^{-0.5 - \alpha' t} (1-\beta)^3 e^{B t}\,,\qquad \alpha' = 0.9~\text{GeV}^{-2}, 
\notag\\
h(\beta,\alpha) &= \frac34 \frac{(1-|\beta|)^2-\alpha^2}{(1-|\beta|)^3} \,, 
\qquad \int\limits_{-1+|\beta|}^{1-|\beta|} d\alpha \, h(\beta,\alpha) =1\,.
\label{GK12}
\end{align}
An overall normalization and the value of the slope parameter $B$ is not important for our purposes as it cancels
in the form factor ratios. For definiteness we take $B_{val} =0$~\cite{Kroll:2012sm}.   

\begin{figure}[t]
\centerline{\includegraphics[width=0.75\textwidth]{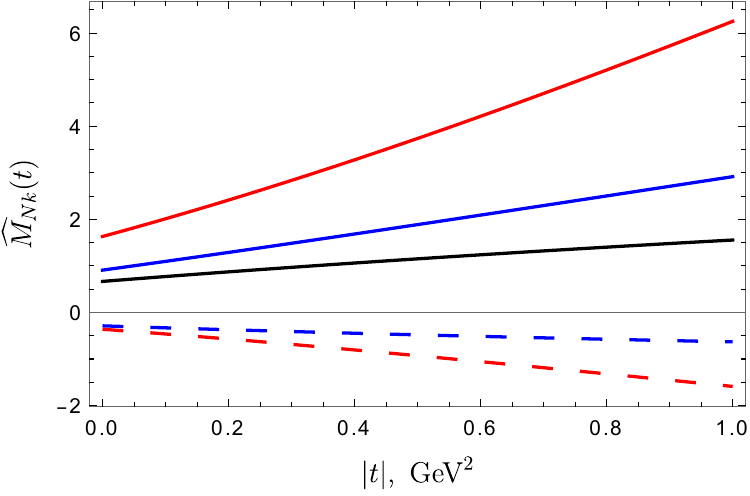}}
\caption{\sf
The Gegenbauer moments $M_{Nk}(t)$ normalized to the nucleon magnetic form factor
$M_{00}(t)$, $\widehat{M}_{Nk}(t) =  M_{Nk}(t)/M_{00}(t)$, for the GPD model in Eq.~\eqref{GK12} with assumed Regge behavior.  
$\widehat{M}_{10}(t)$ is shown by the black solid curve;
$\widehat{M}_{20}(t)$,  $\widehat{M}_{22}(t)$ are shown by the blue solid and blue dashed curves, 
and $\widehat{M_{30}}(t)$,  $\widehat{M_{32}}(t)$ by the red  solid and red dashed curves, respectively.
}
\label{figure:MNkRatios1}
\end{figure}

Gegenbauer moments of the GPD can be expressed directly in terms of integrals of the DD $f^q(\beta,\alpha,t)$, e.g.,
\begin{align}
M_{00}(t) &=  \iint\limits_{|\beta|+|\alpha|\le 1}\!\!d\alpha d\beta\,  f^q(\beta,\alpha,t)\,,
&&
M_{10}(t) =  6\!\! \iint\limits_{|\beta|+|\alpha|\le 1}\!\!d\alpha d\beta\,\beta\,  f^q(\beta,\alpha,t)\,,
\notag\\
M_{20}(t)  &= 30 \iint\limits_{|\beta|+|\alpha|\le 1}d\alpha d\beta\,\beta^2\,  f^q(\beta,\alpha,t)\,,
&&
M_{22}(t) =  \frac32\!\! \iint\limits_{|\beta|+|\alpha|\le 1}\!\!d\alpha d\beta\,  
(5  \alpha^2-1) f^q(\beta,\alpha,t)\,,
\notag\\
M_{30}(t) &=
140\!\! \iint\limits_{|\beta|+|\alpha|\le 1}\!\!d\alpha d\beta\,\beta^3  f^q(\beta,\alpha,t)\,,
&&
M_{32}(t) =
15\!\! \iint\limits_{|\beta|+|\alpha|\le 1}\!\!d\alpha d\beta\,\beta\,(7 \alpha^2-1)   f^q(\beta,\alpha,t)\,,
\end{align}
etc. As a consequence of the assumed Regge behavior, higher moments generally have a weaker $t$-dependence. 
The difference is significant as can be seen from Fig.~\ref{figure:MNkRatios1}, where we plot the ratios
of the Gegenbauer moments $M_{Nk}(t)$ normalized to the nucleon magnetic form factor $M_{00}(t)$ as 
functions of the momentum transfer. 

\begin{figure}[t]
\centerline{\includegraphics[width=0.71\textwidth]{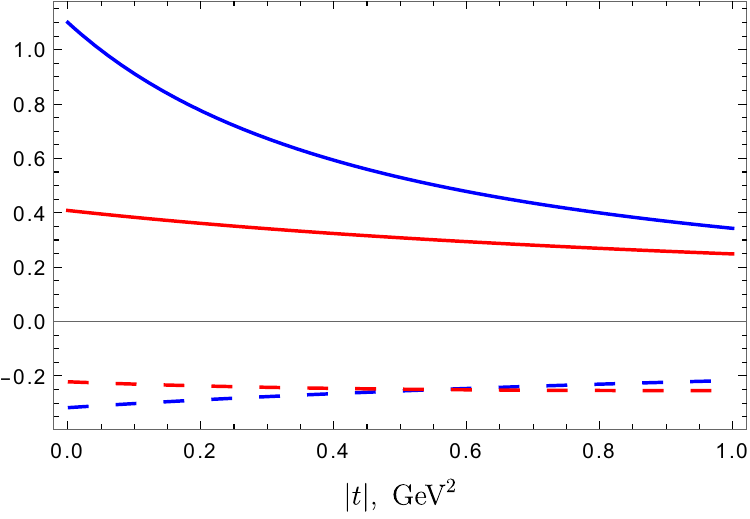}}
\caption{\sf
The form factor ratios appearing in the power corrections to the Gegenbauer moments for $N=2$ and $N=3$, 
Eq.~\eqref{eta=0};
$M_{00}/M_{20}$ and $M_{22}/M_{20}$ are shown by the blue solid and blue dashed curves, 
$M_{10}/M_{30}$ and $M_{32}/M_{30}$ by the red  solid and red dashed curves, respectively.
}
\label{figure:MNkRatios2}
\end{figure}

As an illustration, we consider the zero asymmetry limit, $\eta=0$, in which case the expressions become 
considerably simpler,
\begin{align}
\mathcal{M}^{(P)}_2 &= 
M_{20}(t)\left[ 1
- \frac{1}{18}\frac{v^2P^2}{P_v^2} 
- 3 \frac{v^2t}{P_v^2}  \frac{M_{00}(t)}{M_{20}(t)}
- \frac{1}{18}\frac{v^2t}{P_v^2} \frac{M_{22}(t)}{M_{20}(t)}\right],
\notag\\
\mathcal{M}^{(P)}_3 &= M_{30}(t) 
\left[1 
-  \frac{3}{16} \frac{v^2P^2}{P_v^2} 
- \frac{25}{6} \frac{v^2 t}{P_v^2} \frac{M_{10}(t)}{M_{30}(t)}
- \frac{1}{16} \frac{v^2 t}{P_v^2}  \frac{M_{32}(t)}{M_{30}(t)}\right]. 
\label{eta=0}
\end{align}
The form factor ratios appearing in these expressions are plotted in Fig.~\ref{figure:MNkRatios2}.
One sees that $M_{22}/M_{20} < M_{00}/M_{20}$ and $M_{32}/M_{30} < M_{00}/M_{30}$ and the smaller ratios
also enter with small coefficients. This feature is welcome since it implies that the potential
lattice determination of the GPDs in the zero skewedness limit is only marginally affected by the 
contributions of power corrections involving higher-order terms $\mathcal{O}(\xi^2)$ in the moments 
expansion \eqref{moexp}. The nucleon mass corrections that enter through $P^2 =m_N^2-t/4$ are small
as well. Consequently, the power corrections are driven primarily by the second terms in \eqref{eta=0}, 
which involve  the form factor ratios $M_{00}/M_{20}$ and  $M_{00}/M_{30}$, respectively. They are mitigated somewhat
by the fact that these ratios decrease with the momentum transfer, as illustrated in  Fig.~\ref{figure:MNkRatios2}.
 Combining all factors, for a realistic value $|P_v|/|v| = 2~\text{GeV}$ and $|t|= 1~\text{GeV}^2$ 
one obtains a power correction $\simeq 24\%$  for $\mathcal{M}^{(P)}_2$ and  $\simeq 20\%$ for $\mathcal{M}^{(P)}_3$, respectively.     
A significant part of the kinematic power correction is due to the twist-three contributions, cf. \eqref{M20t2t3}, \eqref{M30t2t3t4}.
The corrections are smaller for  $\mathcal{M}^{(v)}_N$, but this amplitude is more difficult to study on the lattice. 

\begin{figure}[t]
\centerline{\includegraphics[width=0.49\textwidth]{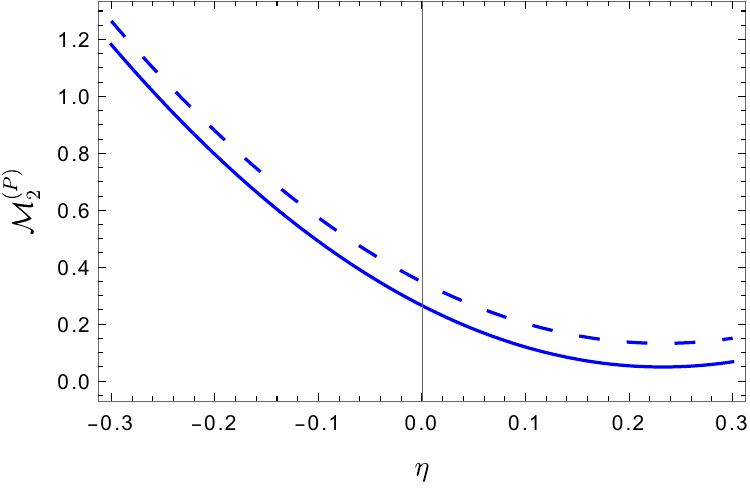}~\includegraphics[width=0.49\textwidth]{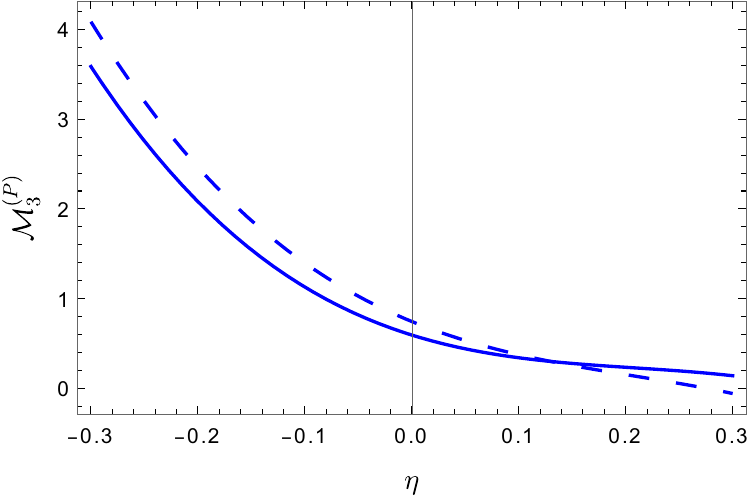}}
\caption{\sf
The Gegenbauer moments $\mathcal{M}^{(P)}_2$ (left panel) and  $\mathcal{M}^{(P)}_3$ (right panel) \eqref{SDE}  
for the GPD model in Eq.~\eqref{GK12} for fixed values  $|t| = 1~\text{GeV}^2$ and $|P_v|/|v| = 2~\text{GeV}$
as functions of the lattice asymmetry parameter \eqref{eta}.
The results with power corrections and without them are shown by the solid and dashed curves, respectively.
}
\label{figure:MP2MP3}
\end{figure}

Kinematic corrections to the quasi-GPD moments at nonzero asymmetry are qualitatively similar. 
This is illustrated in Fig.~\ref{figure:MP2MP3} where we plot $\mathcal{M}^{(P)}_2$ (left panel) and  $\mathcal{M}^{(P)}_3$ (right panel)  
as functions of the lattice asymmetry parameter $\eta$ for fixed values $|t| = 1~\text{GeV}^2$ and $|P_v|/|v| = 2~\text{GeV}$.
The results with and without power corrections are shown by the solid and dashed curves, respectively.

               \section{Conclusions}\label{sec:conclusions}


We have calculated the ``kinematic'' corrections $v^2 t/P_v^2$ and $v^2 m_N^2/P_v^2$ to the short distance 
expansion of gauge-invariant nonlocal quark-antiquark operators sandwiched 
between nucleon states with different momenta.
These matrix elements can be calculated in lattice QCD  and, at leading twist, 
expressed in terms of moments of the generalized parton distributions (GPDs).
The kinematic corrections  turn out to be significant for a realistic lattice QCD setup:
For $|t| = 1~\text{GeV}^2$ and $|P_v|/|v| = 2~\text{GeV}$ we obtain a correction term of the order of 
20-25\% for the $\mathcal{M}^{(P)}$ invariant amplitude \eqref{Vstructures} for zero asymmetry $\eta=0$, for a realistic GPD model \eqref{GK12}.
The dependence on the asymmetry parameter appears to be rather mild, see Fig.~\ref{figure:MP2MP3}.  

From the theory point of view, importance of kinematic power corrections is that they restore the symmetry 
with respect to the translations along the quark-antiquark separation axis, which is broken by the 
twist-expansion~ \cite{Braun:2011zr,Braun:2011dg,Braun:2012bg,Braun:2012hq,Braun:2014sta} . This breaking and its restoration at twist-three 
level for quasi/pseudo-GPDs, $\mathcal{O}(1/|P_v|)$, is demonstrated explicitly in Ref.~\cite{Braun:2023alc}. 
With the expressions obtained in this work, the translation symmetry is restored to the $\mathcal{O}(1/|P_v|^2)$ accuracy.
  
Our results in the final form are presented for the specific choice of quark positions, with the quark field at the origin. This choice 
is favored in lattice calculations~\cite{Bhattacharya:2022aob} and it also leads to somewhat simpler expressions for power corrections 
as compared to the general case. The operator-level results collected in Sec.~\ref{sec:LROPE} are, however,  valid for arbitrary 
quark positions, so that other choices can be worked out starting from these expressions with little effort.

In addition to the kinematic corrections considered in this work, quasi/pseudo-GPD also involve dynamical power corrections 
$\mathcal{O}(v^2\Lambda_{\rm QCD}^2/P_v^2)$ due to the contributions of ``genuine'' twist-four quark-antiquark-gluon operators. 
These corrections were considered in Ref.~\cite{Braun:2025xlp} using renormalon-based techniques. They are separate 
from the ones considered in this work, and there is no double counting.  

Note that the transverse spatial position of partons in the target is Fourier
conjugate to the momentum transfer $t$ in the scattering process.
Thus in order to probe the three-dimensional image of the proton one has to
be able to access sufficiently large values $|t| \gg \Lambda_{\rm QCD}^2$, since $1/\Lambda_{\rm QCD}$ is the nucleon size. 
Under this condition, dynamical power corrections should be less important as compared to the kinematic ones.

\section*{Acknowledgments}
\addcontentsline{toc}{section}{Acknowledgments}

This study was supported by Deutsche Forschungsgemeinschaft (DFG) through the Research Unit FOR 2926, ``Next Generation pQCD for
Hadron Structure: Preparing for the EIC'', project number 40824754. In addition, 
H.-Y.J. gratefully acknowledges support from the National Natural Science Foundation of China with Grant No. 12405114.


\appendix
\addcontentsline{toc}{section}{Appendices}
\renewcommand{\theequation}{\Alph{section}.\arabic{equation}}
\renewcommand{\thesection}{{\Alph{section}}}
\renewcommand{\thetable}{\Alph{table}}
\setcounter{section}{0} \setcounter{table}{0}
\section*{Appendices}

\section{Coefficient functions of kinematic operators}\label{app:A}

The coefficient functions of the kinematic operators in \eqref{eq:victory} are given by~\cite{Braun:2011dg} 
\begin{align}
{\Psi}^{(1)}_{Nk}(z_1,z_2,z_3) &= \Big[(S_+^{111})^k{\Psi}^{(1)}_N\Big](z_1,z_2,z_3)\,,
\notag\\
{\Psi}^{(2)}_{Nk}(z_1,z_2,z_3) &= \Big[(S_+^{1\frac32 \frac12})^k{\Psi}^{(2)}_N\Big](z_1,z_2,z_3)\,,
\end{align}
where $S_+^{j_1,\ldots, j_n}$ are the generators of special conformal transformations acting on
products of fields at different light-cone positions (light-ray operators). They are
given by the sum of the generators acting on the field coordinates, e.g.
\begin{eqnarray}
  S^{(j_1,j_2)}_{+} &=& S^{(j_1)}_{+} + S^{(j_2)}_{+} = z_1^2\partial_{z_1}+z_2^2\partial_{z_2}+2j_1 z_1+2j_2 z_2\,,
\nonumber\\
  S^{(j_1,j_2,j_3)}_{+} &=&
 z_1^2\partial_{z_1}+z_2^2\partial_{z_2}+z_3^2\partial_{z_3}+2j_1 z_1+2j_2 z_2+2j_3 z_3\,.
\label{Splus}
\end{eqnarray}
Also
\begin{align}
\Psi_N^{(1)}(z_1,z_2,z_3)&=
4a_N\left[\int_0^1d\alpha\,\bar\alpha\int_0^1d\beta\,\bar\beta\,
(z_{12}^\alpha-z_{32}^\beta)^{N-1}
-\frac1{N+1}
\int_0^1d\alpha\,\alpha\bar\alpha\,(z_{12}^\alpha-z_3)^{N-1}
\right],
\notag\\
\Psi_N^{(2)}(z_1,z_2,z_3)&=
 -4a_N\int_0^1d\alpha\,\bar\alpha\int_0^1d\beta\,\left(\beta+\frac1{N+1}\alpha\right)
(z_{12}^\alpha-z_{32}^\beta)^{N-1}\,,
\label{WaveFunctions}
\end{align}
where
\begin{align}
a_N=&\frac18(N+3)(N+2)(N+1)N\,.
\end{align}
The functions ${\Psi}^{(1)}_{Nk}$ and ${\Psi}^{(2)}_{Nk}$ 
are mutually orthogonal, and also orthogonal to the coefficient functions of the other existing multiplicatively renormalizable operators.

The following identities are very powerful:
\begin{eqnarray}
 \int_0^1 du\, u\, \Big[(S_+^{(1,\frac32,\frac12)})^k\Psi^{(2)}_N\Big](z_1,z_{21}^u,z_2)
&=& (S_+^{(2,1)})^k   \int_0^1 du\, u\,\Psi^{(2)}_N(z_1,z_{21}^u,z_2)\,,
\nonumber\\
  \int_0^1 du\, \bar u\, \Big[(S_+^{(1,\frac32,\frac12)})^k\Psi^{(2)}_N\Big](z_1,z_{21}^u,z_2)
&=& (S_+^{(\frac32,\frac32)})^k  \int_0^1 du\, \bar u\,\Psi^{(2)}_N(z_1,z_{21}^u,z_2)\,,
\nonumber\\
  \int_0^1 du\, \Big[(S_+^{(1,1,1)})^k\Psi^{(1)}_N\Big](z_1,z_{21}^u,z_2)
&=& (S_+^{(\frac32,\frac32)})^k  \int_0^1 du\,\Psi^{(1)}_N(z_1,z_{21}^u,z_2)\,.
\label{eq:ident}
\end{eqnarray}
Unfortunately, e.g.,
\begin{align}
  \int_0^1 du\,u\, \Big[(S_+^{(1,1,1)})^k\Psi^{(1)}_N\Big](z_1,z_{21}^u,z_2)
\end{align}
cannot be simplified in a similar manner.  
This last integral appears in the OPE for the product of two electromagnetic currents in separate contributions of the
handbag and gluon emission diagrams, but cancels in their sum~\cite{Braun:2011dg}. This cancellation is crucial to obtain
a relatively simple final expression that can be cast in a nonlocal form.

\section{The first few moments}\label{app:B}

In this appendix, we present our results for the power corrections for the lowest Gegenbauer moments
$N\le 3$ for the eleven invariant amplitudes defined in Eqs.~\eqref{Vstructures}, \eqref{Astructures}.
As already mentioned in Sec.~\ref{sec:defqGPD}, this basis is overcomplete. The three structures
in \eqref{Astructures} which involve contributions of axial-vector GPDs and enter at twist-three level, 
can be eliminated at the cost of adding one new structure $\bar u(p') \sigma_{\mu\nu}v^\nu u(p)$. 
If one wants to separate different amplitudes using  projection operators, 
this redundancy has to be taken into account.\\

\noindent\underline{$\bullet$~~$\mathcal{M}^{(v)}_N(t)$}
\begin{align}
 \mathcal{M}^{(v)}_0 &= M_{00}\,,
\qquad\qquad
 \mathcal{M}^{(v)}_1  =   M_{10} - 6\eta M_{00}\,,
\notag\\
 \mathcal{M}^{(v)}_2 &= M_{20} -10\eta M_{10} + 36 \eta^2M_{00}
+4\eta^2 M_{22}
\notag\\&\quad
- \frac16 \frac{v^2P^2}{P_v^2} M_{20}
- \frac16 \frac{v^2t }{P_v^2}\Big[ 9M_{00}+M_{22}\Big],
\notag\\
 \mathcal{M}^{(v)}_3 &=
M_{30} -14\eta M_{20} + 4\eta^2 \big(20M_{10}\!+\!M_{32}\big) 
- 56\eta^3 \big(4 M_{00}\!+\!M_{22}\big) 
\notag\\&\quad
-\frac{1}{24} \frac{v^2P^2}{P_v^2}\Big[9 M_{30} - 56\eta M_{20}\Big]
\notag\\&\quad
-\frac{1}{24} \frac{v^2t}{P_v^2}
\Big[60 M_{10}+ 3 M_{32}-  56 \eta\big(9 M_{00}+ M_{22}\big)\Big].
\end{align}

\vspace*{0.2cm}

\noindent\underline{$\bullet$~~$\mathcal{M}^{(P)}_N(t)$}
\begin{align}
 \mathcal{M}^{(P)}_0 &= M_{00}\,,
\qquad\qquad
 \mathcal{M}^{(P)}_1 = M_{10} - 6\eta M_{00}\,,
\notag\\
\mathcal{M}^{(P)}_2 &= 
M_{20} -10\eta M_{10}  + 36 \eta^2 M_{00}+4 \eta^2 M_{22}
\notag\\&\quad
- \frac{1}{18}\frac{v^2P^2}{P_v^2} M_{20} 
- \frac{1}{18}\frac{v^2t }{P_v^2} \Big[54 M_{00} + M_{22}\Big]\,,  
\notag\\
\mathcal{M}^{(P)}_3 &= M_{30} - 14\eta M_{20} +  4\eta^2\big(20 M_{10}\!+\!M_{32}\big)
 -56\eta^3\big(4M_{00}\!+\!M_{22}\big)
\notag\\&\quad
-  \frac{1}{144} \frac{v^2P^2}{P_v^2} 
\Big[
 27 M_{30}  -  112 \eta M_{20}
\Big]
\notag\\&\quad
- \frac{1}{144} \frac{v^2 t}{P_v^2}
\Big[
 600 M_{10}+9 M_{32} - 112\eta\big(54  M_{00} + M_{22} \big)
\Big]. 
\end{align}

\vspace*{0.2cm}

\noindent\underline{$\bullet$~~$\mathcal{M}^{(\Delta)}_N(t)$}
\begin{align}
 \mathcal{M}^{(\Delta)}_0 &= 0\,, \qquad\qquad  \mathcal{M}^{(\Delta)}_1 =0\,,
\notag\\
 \mathcal{M}^{(\Delta)}_2 &= \frac13\eta \big(9 M_{00} - 4 M_{22}\big),
\notag\\
 \mathcal{M}^{(\Delta)}_3 &=  \frac13 \eta \Big[ 10 M_{10} -3M_{32} -126 \eta M_{00} + 56 \eta M_{22}\Big]
+ \mathcal{O}(\text{twist-five}).
\end{align}
\vspace*{0.2cm}

\noindent\underline{$\bullet$~~$\mathcal{M}^{(\gamma)}_N(t)$}
\begin{align}
 \mathcal{M}^{(\gamma)}_0 &= M_{00}\,,
\qquad\qquad
 \mathcal{M}^{(\gamma)}_1 = \frac12 M_{10} - 6\eta M_{00}\,,
\notag\\
 \mathcal{M}^{(\gamma)}_2 &= \frac13 \Big[ M_{20}  -15\eta M_{10} + 126 \eta^2 M_{00} + 4\eta^2 M_{22}\Big]
\notag\\&\quad
- \frac{1}{18} \frac{v^2P^2}{P_v^2}M_{20} 
- \frac{1}{18} \frac{v^2t}{P_v^2}\Big[54M_{00}+M_{22}\Big] 
+ \mathcal{O}(\text{twist-five})\,,
\notag\\
 \mathcal{M}^{(\gamma)}_3 &= \frac{1}{12}
\Big[3 M_{30} - 56\eta M_{20} +  4 \eta^2 \big(3M_{32}+130 M_{10}\big) - 112 \eta^3\big(33 M_{00}+2 M_{22}\big)\Big]
\notag\\&\quad
- \frac{1}{288}\frac{v^2P^2}{P_v^2} \Big[27 M_{30}- 224\eta M_{20}\Big]
\notag\\&\quad
- \frac{1}{288}\frac{v^2t}{P_v^2}
\Big[600 M_{10} + 9 M_{32}  - 224 \eta\big(54 M_{00} + M_{22}\big) \Big]
+ \mathcal{O}(\text{twist-five})\,.
\end{align}
\vspace*{0.2cm}

\noindent\underline{$\bullet$~~$\mathcal{E}^{(v)}_N(t)$}
\begin{align}
 \mathcal{E}^{(v)}_0 &= E_{00}\,,
\notag\\
\mathcal{E}^{(v)}_1 &=  E_{10} -6\eta E_{00} + 4\eta^2 E_{12} 
- \frac14 \frac{v^2}{P_v^2} \Big[P^2 E_{10} + t E_{12} - m_N^2 M_{10}\Big],
\notag\\
\mathcal{E}^{(v)}_2 &= E_{20}-10\eta E_{10}+ 36 \eta^2 E_{00}+4 \eta^2E_{22} - 40 \eta^3 E_{12}
\notag\\&\quad
- \frac12 \frac{v^2P^2}{P_v^2} \Big[E_{20}- 5\eta E_{10}\Big]
- \frac16 \frac{v^2t}{P_v^2} \Big[ 9 E_{00} - 15\eta E_{12}+ E_{22}\Big]
\notag\\&\quad
+ \frac16 \frac{v^2m_N^2}{P_v^2}\Big[ 2 M_{20}-15\eta M_{10}\Big],
\notag\\
\mathcal{E}^{(v)}_3 &= E_{30} -14 \eta E_{20} + 4\eta^2\big(20 E_{10}+ E_{32}\big) 
-56 \eta^3\big(4 E_{00}+E_{22}\big) 
+ 16 \eta^4 \big(20 E_{12}+E_{34}\big)
\notag\\&\quad
- \frac{1}{108}\frac{v^2P^2}{P_v^2}\Big[81 E_{30}-756 \eta E_{20} + \eta^2 \big(2410 E_{10} + 54 E_{32}\big)\Big]
\notag\\&\quad
- \frac{1}{72}\frac{v^2t}{P_v^2}\Big[
9 E_{32} + 180 E_{10} - 168\eta \big(9 E_{00}+ E_{22}\big) + 4 \eta^2 \big(415 E_{12} + 54 E_{34}\big)\Big] 
\notag\\&\quad
+ \frac{1}{216}\frac{v^2m_N^2}{P_v^2}
\Big[81 M_{30}-1008 \eta M_{20} + 4\eta^2\big(1205 M_{10} + 27 M_{32}\big)\Big].
\end{align}
\vspace*{0.2cm}

\noindent\underline{$\bullet$~~$\mathcal{E}^{(P)}_N(t)$}
\begin{align}
\mathcal{E}_0^{(P)} & = E_{00} \,, \qquad\qquad
\mathcal{E}_1^{(P)}   = E_{10} - 6 \eta  E_{00} + 4 \eta^2 E_{12} \,, 
\notag\\
\mathcal{E}_2^{(P)} & = E_{20} - 10 \eta  E_{10} + 4 \eta^2 \left(9 E_{00}+E_{22}\right) - 40 \eta^3 E_{12} 
\notag \\& \quad 
- \frac16 \frac{v^2P^2}{P_v^2} E_{20} 
- \frac{1}{18}\frac{v^2t}{P_v^2} \left( 54 E_{00} + E_{22} \right) + \frac{m_N^2 v^2}{9 P_v^2 } M_{20} \,, 
\notag\\
\mathcal{E}_3^{(P)} & =  E_{30} - 14 \eta  E_{20} + 4 \eta^2 \left(20 E_{10}+E_{32}\right) - 56 \eta^3 \left(4 E_{00}+E_{22}\right) + 16 \eta^4 \left(20 E_{12}+E_{34}\right) 
\nonumber \\
& \quad + \frac{1}{432}\frac{v^2m_N^2}{P_v^2} \Big[ 81 M_{30} - 672 \eta  M_{20} + 4 \eta^2 \left(190 M_{10} + 27 M_{32}\right) \Big] 
\nonumber \\
& \quad - \frac{1}{216}\frac{v^2P^2}{ P_v^2} \Big[ 81 E_{30} - 504 \eta  E_{20} + 2 \eta^2 \left( 190 E_{10} + 27 E_{32} \right ) \Big] 
\nonumber \\
& \quad - \frac{1}{144}\frac{v^2t }{P_v^2} \Big[ 9 E_{32} + 600 E_{10} - 112 \eta \left(54 E_{00}+ E_{22}\right) + 8\eta^2\left(85 E_{12} + 27 E_{34}\right) \Big] \,, 
\end{align}
\vspace*{0.2cm}

\noindent\underline{$\bullet$~~$\mathcal{E}^{(\Delta)}_N(t)$}
\begin{align}
\mathcal{E}_0^{(\Delta)} & = 0 \,, \qquad\qquad
\mathcal{E}_1^{(\Delta)}  = - 2 \eta  E_{12} \,, 
\notag\\
\mathcal{E}_2^{(\Delta)} & = \frac{1}{3} \eta  \left(9 E_{00} - 4 E_{22}\right) + 20 \eta^2 E_{12} \,, 
\notag\\
\mathcal{E}_3^{(\Delta)} & = \frac{1}{3} \eta \left(10 E_{10} - 3 E_{32}\right) 
+ \frac{14}{3} \eta^2 \left(4 E_{22} - 9 E_{00}\right) - 8 \eta^3 \left(20 E_{12} + E_{34}\right) 
\notag \\
& \quad - \frac{1}{216}\frac{v^2m_N^2}{P_v^2} \eta\left(190 M_{10}+27 M_{32}\right) 
+ \frac{1}{216}\frac{v^2P^2}{P_v^2} \eta \left( 190 E_{10} + 27 E_{32} \right) 
\notag \\
& \quad + \frac{1}{36}\frac{v^2t}{P_v^2} \eta \left( 85 E_{12} + 27 E_{34} \right)
+ \mathcal{O}(\text{twist-five}) \,.
\end{align}

\noindent\underline{$\bullet$~~$\widetilde{\mathcal{H}}^{(\gamma)}_N(t)$}
\begin{align}
\widetilde{\mathcal{H}}_{0}^{(\gamma)} & = 0 \,, \qquad\qquad \widetilde{\mathcal{H}}_{1}^{(\gamma)} = 3 \widetilde{H}_{00} \,, \nonumber \\ 
\widetilde{\mathcal{H}}_{2}^{(\gamma)} & = \frac{5}{3} \widetilde{H}_{10} - 30 \eta \widetilde{H}_{00} \,, \nonumber \\
\widetilde{\mathcal{H}}_{3}^{(\gamma)} & = \frac{7}{6} \left[ \widetilde{H}_{20} - 20 \eta \widetilde{H}_{10} + 4 \eta^2 \left(54 \widetilde{H}_{00} + \widetilde{H}_{22} \right) - \frac{v^2 t}{6 P_v^2} \left(54 \widetilde{H}_{00} + \widetilde{H}_{22}\right) - \frac{v^2 P^2}{6 P_v^2} \widetilde{H}_{20}\right] \nonumber \\
& \quad + \mathcal{O}(\text{twist-five}) \,.
\end{align}

\noindent\underline{$\bullet$~~$\widetilde{\mathcal{H}}^{(\tilde{\Delta})}_N(t)$}
\begin{align}
\widetilde{\mathcal{H}}_{0}^{(\Delta)} & = 0 \,, \qquad\qquad \widetilde{\mathcal{H}}_{1}^{(\Delta)} = 0 \,, \qquad\qquad \widetilde{\mathcal{H}}_{2}^{(\Delta)} = -\frac{5}{6} \widetilde{H}_{10} \,, \nonumber \\
\widetilde{\mathcal{H}}_{3}^{(\Delta)} & = \frac{7}{6} \left(10 \eta  \widetilde{H}_{10} - \widetilde{H}_{20}\right) \,.
\end{align}

\noindent\underline{$\bullet$~~$\widetilde{\mathcal{E}}^{(\tilde{\Delta})}_N(t)$}
\begin{align}
\widetilde{\mathcal{E}}_{0}^{(\Delta)} & = 0 \,, \qquad\qquad \widetilde{\mathcal{E}}_{1}^{(\Delta)} = 0 \,, \qquad\qquad
\widetilde{\mathcal{E}}_{2}^{(\Delta)} = -\frac{5}{6} \widetilde{E}_{10} \,, \nonumber \\
\widetilde{\mathcal{E}}_{3}^{(\Delta)} & = - \frac{7}{6} \left[ \widetilde{E}_{20} - 10 \eta  \widetilde{E}_{10} + \frac{5}{12} \frac{v^2 t}{\eta P_v^2} \widetilde{E}_{10} + \frac{5}{3} \frac{v^2 m_N^2}{\eta P_v^2} \widetilde{H}_{10} \right] + \mathcal{O}(\text{twist-five})\,.
\end{align}
The addenda $ \mathcal{O}(\text{twist-five})$ serves to remind that these amplitudes can receive power corrections 
$t/P^2_v$, $m_N^2/P^2_v$ from the contributions of twist-five operators which are not taken into account in this work. 


               \section{Axial-vector quasi/pseudo GPDs}\label{app:C}


In this appendix we present the results for the twist expansion of nucleon matrix elements  of 
the axial-vector nonlocal quark-antiquark  operator
$\bar q (z_1v)\gamma^\mu\gamma_5 q(z_2v)$, retaining kinematic contributions only. 
The operator-level expressions are similar to the vector case \eqref{t23}:
\begin{align}
 [ \bar q(z_1v) \gamma^\mu\gamma_5 q(z_2v)]_{t2} &= \partial^\mu\int_0^1\!du\, 
[\bar q(z_1uv)\slashed{v}\gamma_5q(z_2uv)]_{lt}\,,
\notag\\
 [ \bar q(z_1v) \gamma^\mu\gamma_5 q(z_2v)]_{t3} &=
\frac12 \int_0^1\!udu \!\int_{z_2}^{z_1}\frac{dw}{z_{12}} 
\biggl\{
\Big[(vd)\partial^\mu - (v\partial) d^\mu + v^\mu (d\partial)+ \ln u\, \partial^\mu  v^2 (d \partial) \Big]
\notag\\&\hspace*{3cm}
\times \Big(z_1 [\bar q(z_1uv)\slashed{v}\gamma_5q(wuv)]_{lt} + z_2 [\bar q(wuv)\slashed{v}\gamma_5q(z_2uv)]_{lt}\Big)
\notag\\&\quad
+i\epsilon^{\rho\nu\sigma\mu} v_\rho \partial_\sigma \nabla_\nu  
\Big (z_1 [\bar q(z_1uv)\slashed{v}q(wuv)]_{lt} - z_2 
[\bar q(wuv)\slashed{v}q(z_2uv)]_{lt}\Big)
\biggr\},
\notag\\
  [ \bar q(z_1v) \gamma^\mu q(z_2v)]_{t4} &= 2 v^\mu [\widetilde A(v;z_1,z_2)]_{lt} 
  + 2 v^2 \partial^\mu [\widetilde B (v;z_1,z_2)]_{lt}\,,
\label{t234axial}
\end{align}
where
\begin{align}\label{A1axial}
\widetilde A(n;z_1,z_2)&=\frac14\int_0^1du \,\biggl\{ u^2\ln u\,
z_1z_2\, \nabla^2\, \big[ \bar q(uz_1n) \slashed{n}\gamma_5 q(uz_2un)\big]
\notag\\
&\quad+\left(z_2\partial_{z_2}-\frac{z_1}{z_{12}}-\ln u\, z_2\partial_{z_2}^2 z_{12}\right) R_2(uz_1n,uz_2n)
\notag\\&\quad
-\left(z_1\partial_{z_1}-\frac{z_2}{z_{21}}-\ln u\, z_1\partial_{z_1}^2 z_{21}\right) \bar R_2(uz_1n,uz_2n)
\biggr\}\,,
\\ 
\label{B1axial}
\widetilde B(n;z_1,z_2) &=\frac18\int_0^1\frac{du}{u^2} \,\biggl\{u^2(1\!-\!u^2\!+\!u^2\ln u)\,
z_1z_2\,\nabla^2\, \big[ \bar q(uz_1n) \slashed{n}\gamma_5 q(uz_2un)\big]
\notag\\&\quad
 -\left[
(1-u^2)\left(z_2\partial_{z_2}-\frac{z_1}{z_{12}}\right)+(1\!-\!u^2\!+\!u^2\ln u)\,
z_2\partial_{z_2}^2 z_{12}\right] \widetilde R_2(uz_1n,uz_2n)
\notag\\&\quad
+\left[
(1-u^2)\left(z_1\partial_{z_1}-\frac{z_2}{z_{21}}\right)+(1\!-\!u^2\!+\!u^2\ln u)\,
z_1\partial_{z_1}^2 z_{21}\right] \bar{\widetilde R}_2(uz_1n,uz_2n)
\biggl\}
\notag\\&\quad
+ \frac18 \int_0^1\frac{du}{u^2} \,\biggl[  \widetilde R_1(uz_1n,uz_2n) - \bar{\widetilde  R}_1(uz_1n,uz_2n)\biggr].
\end{align}
The $\widetilde R$-operators are defined as in Eq.~\eqref{RbarR} with the replacement of $Q_1$ and $Q_2$
by their axial-vector counterparts 
\begin{align}
\widetilde Q_2(z_1n, w n,z_2n)  - \widetilde Q_1(z_1n,w n,z_2n)  &=  
\bar q(z_1n)  \big[ g \widetilde F_{+\mu}(wn) - ig F_{+\mu}(wn)\gamma_5\big]\gamma^\mu q(z_2n)\,,
\notag\\
 \bar{\widetilde Q}_i(z_1n,wn,z_2n)& =(\widetilde Q_i(z_2n,wn,z_1n))^\dagger.
\end{align}

The contributions of the axial-vector GPDs to the nucleon matrix element of the axial-vector operator can be written as 
\begin{align}
 \langle p'|Q^{(\gamma_\mu\gamma_5)}|p\rangle^A &=   
 \bar u(p')\slashed{v}\gamma_5 u(p)
\biggl\{\frac{v^\mu}{v^2} \,\widetilde{\mathfrak{H}}^{(v)}
+  \left(\frac{P^\mu}{(Pv)} -  \frac{v^\mu}{v^2}\right) \widetilde{\mathfrak{H}}^{(P)}
+ \left(\frac{\Delta^\mu_\perp}{(Pv)}\right) \widetilde{\mathfrak{H}}^{(\Delta)}\biggr\}
\notag\\&\quad
+ \bar u(p')\left(\gamma_\mu- \frac{P_\mu\slashed{v}}{(Pv)}\right)\gamma_5 u(p) \widetilde{\mathfrak{H}}^{(\gamma)} 
\notag\\&\quad
  + \frac{(\Delta v)}{2 m_N}\bar u(p')\gamma_5u(p)
\biggl\{ \frac{v^\mu}{v^2} \, \widetilde{\mathfrak{E}}^{(v)}
+  \left(\frac{P^\mu}{(Pv)} - \frac{v^\mu}{v^2}\right) \widetilde{\mathfrak{E}}^{(P)} 
+ \left(\frac{\Delta^\mu_\perp}{(Pv)}\right) 
\widetilde{\mathfrak{E}}^{(\Delta)}\biggr\}.
\label{AxialVstructures}
\end{align}
At twist-three level also the vector GPDs $M=H+E$ and $E$ enter the game. 
We have chosen to leave these contributions in the form how they appear in the calculation, 
\begin{align}
 \langle p'|Q^{(\gamma_\mu\gamma_5)}|p\rangle^V &=   
 \frac{i}{2}\epsilon^{\rho\nu\sigma\mu} \frac{v_\rho \Delta_\nu}{(Pv)}  \bar u(p') \gamma^\sigma u(p) 
\,{\mathfrak{M}}^{(\gamma)} 
+ \frac{\widetilde{\Delta}^\perp_\mu}{(Pv)} \bar u(p')\slashed{v}u(p) {\mathfrak{M}}^{(\tilde\Delta)}
\notag\\&\quad
+  \frac{\widetilde{\Delta}^\perp_\mu}{(Pv)}  \frac{(\Delta v)}{2 m_N}\bar u(p') u(p) {\mathfrak{E}}^{(\tilde \Delta)}.   
\label{AxialAstructures}
\end{align}
These extra terms can be decomposed in terms of 
the above seven Lorentz/Dirac structures \eqref{AxialVstructures}, with one extra structure
 $\bar u (p') i \sigma_{\mu\nu} v^\nu\gamma_5 u(p)$. For example
\begin{align}
 \frac{i}{2}\epsilon^{\rho\nu\sigma\mu} \frac{v_\rho \Delta_\nu}{(Pv)}  \bar u (p') \gamma^\sigma u(p) 
&= \bar u(p')\left(\gamma_\mu- \frac{P_\mu\slashed{v}}{(Pv)}\right)\gamma_5 u(p)
  -  \frac{m_N}{(Pv)} \bar u (p') i \sigma_{\mu\nu} v^\nu\gamma_5 u(p)\,. 
\end{align}
This rewriting is straightforward but leads to longer expressions.

\subsection{Twist-two}
We write the results as
\begin{align}
 (\widetilde{\mathfrak F})_{t2} &= 
 \sum_{N=0}^\infty 2(2N+3) \sum_{k=0}^\infty \frac{(iz P_v)^{N+k}}{k!}
\frac{\Gamma(N+k+1)}{\Gamma(2N+k+4)}  
(\widetilde{\mathfrak F}_{Nk})_{t2}
\end{align}
where \eqref{AxialVstructures}  
\begin{align}
 \widetilde{\mathfrak F} \in \{  \widetilde{\mathfrak H}^{(v)}, \widetilde{\mathfrak H}^{(P)}, \widetilde{\mathfrak H}^{(\Delta)}, 
\widetilde{\mathfrak H}^{(\gamma)}, \widetilde{\mathfrak E}^{(v)}, \widetilde{\mathfrak E}^{(P)}, \widetilde{\mathfrak E}^{(\Delta)}\}.
\label{TwistTwoSetAxial}
\end{align}
We obtain
\begin{align}
(\widetilde{\mathfrak{H}}^{(v)}_{Nk})_{t2} 
& =  (N+k+1) \sum_{\substack{m=0, \\ \text{even}}}^{N}(-2\eta)^{m+k} \widetilde{H}_{Nm}  
- \frac{v^2}{4P_v^2} \sum_{\substack{m=-2, \\\text{even}}}^{N-2} (-2\eta)^{m+k} \widetilde{\mathbb{H}}_{Nkm}^{(4)} \,, 
\notag\\
(\widetilde{\mathfrak{H}}^{(P)}_{Nk})_{t2} 
& = (N+k+1) \sum_{\substack{m=0, \\\text{even}}}^{N} (-2\eta)^{m+k} \widetilde{H}_{Nm} 
- \frac{v^2}{4P_v^2} \frac{N+k-1}{N+k+1} \sum_{\substack{m=-2, \\ \text{even}}}^{N-2} 
(-2\eta)^{m+k} \widetilde{\mathbb{H}}_{Nkm}^{(4)} \,, 
\notag\\
(\widetilde{\mathfrak{H}}^{(\Delta)}_{Nk})_{t2} 
& = \sum_{\substack{m=0, \\ \text{even}}}^{N} (m+k) (-2\eta )^{m+k-1} \widetilde{H}_{Nm} 
- \frac{v^2}{4P_v^2}\sum_{\substack{m=-2, \\ \text{even}}}^{N-2} \frac{m+k}{N+k+1} (-2\eta)^{m+k-1} \widetilde{\mathbb{H}}_{Nkm}^{(4)} \,, 
\notag\\
(\widetilde{\mathfrak{H}}^{(\gamma)}_{Nk})_{t2} 
& = \sum_{\substack{m=0, \\ \text{even}}}^{N}(-2\eta)^{m+k} \widetilde{H}_{Nm} 
- \frac{v^2}{4P_v^2} \frac{1}{N+k+1} \sum_{\substack{m=-2, \\ \text{even}}}^{N-2} (-2\eta)^{m+k} \widetilde{\mathbb{H}}_{Nkm}^{(4)} \,,
\end{align}
and
\begin{align}
(\widetilde{\mathfrak{E}}^{(v)}_{Nk})_{t2} 
& =  (N+k+1) \sum_{\substack{m=0,\\\text{even}}}^{N} (-2\eta)^{m+k} \widetilde{E}_{Nm} 
- \frac{v^2}{4P_v^2} 
\sum_{\substack{m=-2,\\ \text{even}}}^{N-2}(-2\eta)^{m+k} \widetilde{\mathbb{E}}_{Nkm}^{(4)} \,, 
\notag\\
(\widetilde{\mathfrak{E}}^{(P)}_{Nk})_{t2} 
& = (N+k+1)  \sum_{\substack{m=0,\\\text{even}}}^{N}(-2\eta)^{m+k} \widetilde{E}_{Nm} 
- \frac{v^2}{4P_v^2} \frac{N+k-1}{N+k+1} \sum_{\substack{m=-2,\\ \text{even}}}^{N-2}(-2\eta)^{m+k} \widetilde{\mathbb{E}}_{Nkm}^{(4)} \,, 
\notag\\
(\widetilde{\mathfrak{E}}^{(\Delta)}_{Nk})_{t2} &= 
\sum_{\substack{m=0,\\ \text{even}}}^{N} (m+k+1) (-2\eta )^{m+k-1} \widetilde{E}_{Nm} 
 - \frac{v^2}{4P_v^2}  \sum_{\substack{m=-2,\\ \text{even}}}^{N-2} \frac{m+k+1}{N+k+1}  \big(-2\eta\big)^{m+k-1} \widetilde{\mathbb{E}}_{Nkm}^{(4)}  \,,
\end{align}
where $\widetilde{\mathbb{H}}^{(4)}_{Nkm}$ and $\widetilde{\mathbb{E}}^{(4)}_{Nkm}$ are defined in Eq.~\eqref{tildeMathbbHE}.

\subsection{Twist-three}
Twist-three contributions can be written as
\begin{align}
 (\widetilde{\mathfrak F})_{t2} &= 
 \sum_{N=0}^\infty  \frac{(2N+3)}{(N+1)(N+2)} \sum_{k=0}^\infty \frac{(iz P_v)^{N+k+1}}{k!}\frac{\Gamma(N+k+2)}{\Gamma(2N+k+4)}
(\widetilde{\mathfrak F}_{Nk})_{t2}
\end{align}
with the invariant amplitudes \eqref{AxialVstructures}, \eqref{AxialAstructures}  
\begin{align}
 \widetilde{\mathfrak F} \in \{  \widetilde{\mathfrak H}^{(v)}, \widetilde{\mathfrak H}^{(P)}, \widetilde{\mathfrak H}^{(\Delta)}, 
\widetilde{\mathfrak H}^{(\gamma)}, \widetilde{\mathfrak E}^{(v)}, \widetilde{\mathfrak E}^{(P)}, \widetilde{\mathfrak E}^{(\Delta)},
{\mathfrak{M}}^{(\gamma)},{\mathfrak{M}}^{(\tilde \Delta)}, {\mathfrak{E}}^{(\tilde \Delta)} \}.
\label{TwistThreeSetAxial}
\end{align}
We obtain
\begin{align}
(\widetilde{\mathfrak{H}}^{(v)}_{Nk})_{t3}&= 0\,,
\notag\\
(\widetilde{\mathfrak{H}}^{(P)}_{Nk})_{t3}&=   
\frac{1}{(N+k+2)}\frac{v^2}{P_v^2}\sum_{\substack{m=-2,\\ \text{even}}}^{N-2} (-2\eta)^{m+k+1}  
\biggl\{\widetilde{\mathbb H}^{(4)}_{Nkm} - (N+k) (m+k+2) t \widetilde{H}_{Nm+2}\biggr\},
\notag\\
%
%
(\widetilde{\mathfrak{H}}^{(\Delta)}_{Nk})_{t3}&= 
- \sum_{\substack{m=-2,\\ \text{even}}}^N  (N\!-\!m\!+\!1) (-2\eta)^{m+k} \widetilde{H}_{Nm}
+ \frac{ 1 }{4(N\!+\!k\!+\!2)} \frac{v^2}{P_v^2} \sum_{\substack{m=-2,\\ \text{even}}}^{N-2}\!(N\!-\!m\!-\!1)(-2\eta)^{m+k}
\notag\\&\qquad\qquad  
\times
\biggl\{\frac{N+k}{N\!+\!k\!+\!1}\widetilde{\mathbb H}^{(4)}_{Nkm} \!+ 4 (N\!-\!m) P^2 \tH_{Nm}\biggr\},
\notag\\
(\widetilde{\mathfrak{H}}^{(\gamma)}_{Nk})_{t3}&= 
\sum_{\substack{m=0,\\ \text{even}}}^N (-2\eta)^{m+k+1} \widetilde{H}_{Nm}
-\frac{1}{4(N\!+\!k\!+\!2)}\frac{v^2}{P_v^2}\sum_{\substack{m=-2,\\ \text{even}}}^{N-2}
(-2\eta)^{m+k+1}
\notag\\&\qquad\qquad
\times\biggl\{
\frac{N+k}{N\!+\!k\!+\!1}\widetilde{\mathbb H}^{(4)}_{Nkm} + 4 (m\!+\!k\!+\!2) t \widetilde{H}_{Nm+2}\biggr\}, 
\end{align}
\begin{align}
(\widetilde{\mathfrak{E}}^{(v)}_{Nk})_{t3}&= 0\,,
\notag\\
(\widetilde{\mathfrak{E}}^{(P)}_{Nk})_{t3}&=
\frac{1}{(N+k+2)} \frac{v^2}{P_v^2}  \sum_{\substack{m=-2,\\ \text{even}}}^{N-2} (-2\eta)^{m+k+1} \widetilde{\mathbb E}^{(4)}_{Nm}  
\notag\\&\quad
- \frac{(N+k) }{(N+k+2)} \frac{v^2}{P_v^2} \sum_{\substack{m=-2,\\ \text{even}}}^{N-2}(-2\eta)^{m+k+1}  
\biggl[ (m+k+3) t \tE_{Nm+2} + 4 m_N^2 \widetilde{H}_{Nm+2}\biggr],
\notag\\
(\widetilde{\mathfrak{E}}^{(\Delta)}_{Nk})_{t3}&=
%
%
- \sum_{\substack{m=0,\\ \text{even}}}^N (N-m) (-2\eta)^{m+k}\tE_{Nm}
  +  \frac{1}{4} \frac{N\!+\!k}{N\!+\!k\!+\!1}  \frac{v^2}{P_v^2}  \sum_{\substack{m=-2,\\ \text{even}}}^{N-2}
\frac{N\!-\!m\!-\!2}{N\!+\!k\!+\!2}   (-2\eta)^{m+k} \widetilde{\mathbb E}^{(4)}_{Nm} 
\notag\\&\quad
  +  \frac{1}{(N+k+2)} \frac{v^2}{P_v^2}  \sum_{\substack{m=-2,\\ \text{even}}}^{N-2} 
(N\!-\!m)(N\!-\!m\!-\!1)  (-2\eta)^{m+k} 
 P^2\widetilde E_{Nm} 
\notag\\&\quad
+   4 \frac{v^2}{P_v^2} \sum_{\substack{m=-2,\\ \text{even}}}^{N-2} \frac{(m+k+2)}{(N+k+2)}  (-2\eta)^{m+k}
m_N^2 \widetilde{H}_{Nm+2} \,,
\end{align}
and 
\begin{align}
({\mathfrak{M}}^{(\gamma)})_{t3}  &= 
2\sum_{\substack{m=0,\\ \text{even}}}^N (-2\eta)^{m+k} M_{Nm}
-\frac{1}{2(N+k+1)} \frac{v^2}{P_v^2}  \sum_{\substack{m=-2,\\ \text{even}}}^{N-2}
(-2\eta)^{m+k}  \mathbb M^{(4)}_{Nkm}\,,
\notag\\
({\mathfrak{M}}^{(\tilde \Delta)})_{t3}  &=
- \sum_{\substack{m=0,\\ \text{even}}}^N (N\!-\!m) (-2\eta)^{m+k} M_{Nm}
+\frac{1}{4} \frac{v^2}{P_v^2} \sum_{\substack{m=-2,\\ \text{even}}}^{N-2} 
\frac{N\!-\!m\!-\!2}{N\!+\!k\!+\!1} (-2\eta)^{m+k}\mathbb M^{(4)}_{Nkm} \,,
\notag\\
({\mathfrak{E}}^{(\tilde \Delta)})_{t3}  &=
2 \sum_{\substack{m=0,\\ \text{even}}}^{N+1}(N\!-\!m\!+\!1) (-2\eta)^{m+k-1} E_{Nm}
-\frac{1}{2} \frac{v^2}{P_v^2} \sum_{\substack{m=-2,\\ \text{even}}}^{N-1}
\frac{N\!-\!m\!-\!1}{N\!+\!k\!+\!1} (-2\eta)^{m+k-1} \mathbb E^{(4)}_{Nkm}.
\end{align}

\subsection{Twist-four}
Compared to the vector case, the only modification  is a different expression 
for the matrix element of the kinematic twist-four operator, see \eqref{partialOMEAxial}, \eqref{partialOMEAxial2}.
The resulting expressions for the invariant amplitudes
$
 \widetilde{\mathfrak F} \in \{  \widetilde{\mathfrak H}^{(v)}, \widetilde{\mathfrak H}^{(P)}, \ldots\}
$  \eqref{AxialVstructures}  
take the form
\begin{align}
 (\widetilde{\mathfrak F})_{t4} &= 
 \frac12 \sum_{N=0}^\infty  \frac{(2N+3)}{(N+1)^2(N+2)^2} \sum_{k=0}^\infty \frac{(iz P_v)^{N+k+1}}{k!}
 \frac{\Gamma(N+k+2)}{\Gamma(2N+k+4)}
\biggl[
 (\widetilde{\mathfrak F}^A_{Nk})_{t4}
+ T_{Nk} (\widetilde{\mathfrak F}^B_{Nk})_{t4}\biggr],
\end{align}
where $(\widetilde{\mathfrak F}^A_{Nk})_{t4}$ and  $(\widetilde{\mathfrak F}^B_{Nk})_{t4}$ correspond to the contributions in 
Eqs.~\eqref{A1axial} and \eqref{B1axial}, respectively, $T_{Nk}$ is defined in \eqref{TNk} and  
$S(N,k)$, $\bar S(N,k)$ in \eqref{SbarS}. 
We obtain
\begin{align}
(\widetilde{\mathfrak{H}}^{(v),A}_{Nk})_{t4}&=
  \frac{v^2}{P_v^2} \sum_{\substack{m=-2,\\ \text{even}}}^{N-2}(-2\eta)^{m+k+1} (\partial \widetilde{\mathbb H})_{Nm}\,,
\notag\\
(\widetilde{\mathfrak{H}}^{(v),B}_{Nk})_{t4} &= (\widetilde{\mathfrak{H}}^{(P),B}_{Nk})_{t4} = 
 \frac{v^2}{P_v^2} \sum_{\substack{m=-2,\\ \text{even}}}^{N-2} (N+k) (-2\eta)^{m+k+1}  (\partial \widetilde{\mathbb H})_{Nm}\,,
\notag\\
(\widetilde{\mathfrak{H}}^{(\Delta),B}_{Nk})_{t4} &=
 \frac{v^2}{P_v^2} \sum_{\substack{m=-2,\\ \text{even}}}^{N-2} (m+k+1) (-2\eta)^{m+k}  (\partial \widetilde{\mathbb H})_{Nm}\,,
\notag\\
(\widetilde{\mathfrak{H}}^{(\gamma),B}_{Nk})_{t4} &=
  \frac{v^2}{P_v^2} \sum_{\substack{m=-2,\\ \text{even}}}^{N-2} (-2\eta)^{m+k+1}  (\partial \widetilde{\mathbb H})_{Nm}\,,
\end{align}
and
\begin{align}
(\widetilde{\mathfrak{E}}^{(v),A}_{Nk})_{t4}&=
 \frac{v^2}{P_v^2} \sum_{\substack{m=-2,\\ \text{even}}}^{N-2} (-2\eta)^{m+k+1} (\partial \widetilde{\mathbb E})_{Nm}\,,  
\notag\\
(\widetilde{\mathfrak{E}}^{(v),B}_{Nk})_{t4}&= 
(\widetilde{\mathfrak{H}}^{(P),B}_{Nk})_{t4} =
 \frac{v^2}{P_v^2} \sum_{\substack{m=-2,\\ \text{even}}}^{N-2}  (N+k) (-2\eta)^{m+k+1}  (\partial \widetilde{\mathbb E})_{Nm}\,,
\notag\\
(\widetilde{\mathfrak{E}}^{(\Delta),B}_{Nk})_{t4}&=
 \frac{v^2}{P_v^2} \sum_{\substack{m=-2,\\ \text{even}}}^{N-2} (m+k+2) (-2\eta)^{m+k}  (\partial \widetilde{\mathbb E})_{Nm}\,.
\end{align}
Note that the $A$-structure \eqref{A1axial} only contributes to $(\mathcal{M}^{(v),A}_{Nk})_{t4}$ and $(\mathcal{E}^{(v),A}_{Nk})_{t4}$.

\subsection{The first few moments}
Here, we also present the first four moments for the axial-vector GPDs:

\noindent\underline{$\bullet$~~$\widetilde{\mathfrak{M}}^{(v)}_N(t)$}
\begin{align}
\widetilde{\mathfrak{M}}_0^{(v)} & = \widetilde{H}_{00} \,, \qquad\qquad 
\widetilde{\mathfrak{M}}_1^{(v)} = \widetilde{H}_{10} - 6 \eta  \widetilde{H}_{00}  \,, \nonumber \\
\widetilde{\mathfrak{M}}_2^{(v)} & = \widetilde{H}_{20} - 10 \eta  \widetilde{H}_{10} + 4 \eta^2 \left(9 \widetilde{H}_{00}+\widetilde{H}_{22}\right) - \frac{t v^2}{6 P_v^2} \left(9 \widetilde{H}_{00} + \widetilde{H}_{22} \right) -\frac{P^2 v^2}{6 P_v^2} \widetilde{H}_{20}  \,, \nonumber \\
\widetilde{\mathfrak{M}}_3^{(v)} & = \widetilde{H}_{30} - 14 \eta  \widetilde{H}_{20} + 4 \eta ^2 \left(20 \widetilde{H}_{10}+\widetilde{H}_{32}\right) - 56 \eta^3 \left(4 \widetilde{H}_{00}+\widetilde{H}_{22}\right) \nonumber \\
& \quad + \frac{1}{24}\frac{t v^2}{P_v^2} \left[ 56 \eta  \left(9 \widetilde{H}_{00} + \widetilde{H}_{22}\right) - 60 \widetilde{H}_{10} - 3\widetilde{H}_{32}\right] \nonumber \\
& \quad  + \frac{1}{24} \frac{P^2 v^2}{P_v^2}  \left(56 \eta  \widetilde{H}_{20} - 9 \widetilde{H}_{30}\right) \,.
\end{align}

\noindent\underline{$\bullet$~~$\widetilde{\mathfrak{M}}^{(P)}_N(t)$}
\begin{align}
\widetilde{\mathfrak{M}}_0^{(P)} & = \widetilde{H}_{00} \,, \qquad\qquad
\widetilde{\mathfrak{M}}_1^{(P)} = \widetilde{H}_{10} - 6 \eta  \widetilde{H}_{00} \,, \nonumber \\
\widetilde{\mathfrak{M}}_2^{(P)} & = \widetilde{H}_{20} - 10 \eta  \widetilde{H}_{10} + 4 \eta^2 \left(9 \widetilde{H}_{00}+\widetilde{H}_{22}\right) - \frac{1}{18} \frac{t v^2}{P_v^2} \left( 54 \widetilde{H}_{00} + \widetilde{H}_{22}\right) - \frac{P^2 v^2}{18 P_v^2} \widetilde{H}_{20} \,, \nonumber \\
\widetilde{\mathfrak{M}}_3^{(P)} & = \widetilde{H}_{30}-14 \eta  \widetilde{H}_{20}  + 4 \eta^2 \left(20 \widetilde{H}_{10}+\widetilde{H}_{32}\right) - 56 \eta^3 \left(4 \widetilde{H}_{00}+\widetilde{H}_{22}\right) \nonumber \\
& \quad + \frac{1}{144} \frac{t v^2}{P_v^2} \left[ 112 \eta  \left(54 \widetilde{H}_{00}+\widetilde{H}_{22}\right) - 600 \widetilde{H}_{10} - 9 \widetilde{H}_{32}\right] \nonumber \\
& \quad + \frac{1}{144} \frac{P^2 v^2}{P_v^2} \left( 112 \eta  \widetilde{H}_{20} - 27 \widetilde{H}_{30}\right) \,. 
\end{align}

\noindent\underline{$\bullet$~~$\widetilde{\mathfrak{M}}^{(\Delta)}_N(t)$}
\begin{align}
\widetilde{\mathfrak{M}}_0^{(\Delta )} & = 0 \,, \qquad\qquad
\widetilde{\mathfrak{M}}_1^{(\Delta )} = \frac{3}{2} \widetilde{H}_{00} \,, \qquad\qquad
\widetilde{\mathfrak{M}}_2^{(\Delta )} = \frac{5}{6} \widetilde{H}_{10} -\frac{4}{3} \eta  \left(9 \widetilde{H}_{00}+\widetilde{H}_{22}\right) \,, \nonumber \\
\widetilde{\mathfrak{M}}_3^{(\Delta )} & = \frac{7}{12} \widetilde{H}_{20} - \frac{1}{3} \eta  \left( 25 \widetilde{H}_{10} + 3 \widetilde{H}_{32}\right) + 21 \eta ^2 \left(4 \widetilde{H}_{00}+\widetilde{H}_{22}\right)  \nonumber \\
& \quad - \frac{7}{144} \frac{t v^2}{P_v^2} \left(54 \widetilde{H}_{00}+\widetilde{H}_{22}\right) - \frac{49}{144} \frac{P^2 v^2}{P_v^2} \widetilde{H}_{20} 
+ \mathcal{O}(\text{twist-five}) \,.
\end{align}

\noindent\underline{$\bullet$~~$\widetilde{\mathfrak{M}}^{(\gamma)}_N(t)$}
\begin{align}
\widetilde{\mathfrak{M}}_0^{(\gamma )} & = \widetilde{H}_{00} \,, \qquad\qquad
\widetilde{\mathfrak{M}}_1^{(\gamma )} =\frac{1}{2} \left(6 \widetilde{H}_{00}+\widetilde{H}_{10}\right)-3 \eta  \widetilde{H}_{00} \,, \nonumber \\
\widetilde{\mathfrak{M}}_2^{(\gamma )} & = \frac{1}{3} \left(5 \widetilde{H}_{10}+\widetilde{H}_{20}\right) - \frac{10}{3} \eta  \left(9 \widetilde{H}_{00}+\widetilde{H}_{10}\right)+\frac{4}{3} \eta ^2 \left(9 \widetilde{H}_{00}+\widetilde{H}_{22}\right) \nonumber \\
& \quad - \frac{t v^2}{18 P_v^2} \left(9 \widetilde{H}_{00}+\widetilde{H}_{22}\right) -\frac{P^2 v^2}{18 P_v^2} \widetilde{H}_{20}  \,, \nonumber \\
\widetilde{\mathfrak{M}}_3^{(\gamma )} & = \frac{1}{12} \left(14 \widetilde{H}_{20}+3 \widetilde{H}_{30}\right) - \frac{7}{6} \eta  \left(20 \widetilde{H}_{10}+3 \widetilde{H}_{20}\right) \nonumber \\
& \quad + \frac{1}{3} \eta ^2 \left(756 \widetilde{H}_{00}+60 \widetilde{H}_{10}+14 \widetilde{H}_{22}+3 \widetilde{H}_{32}\right)  - 14 \eta^3 \left(4 \widetilde{H}_{00}+\widetilde{H}_{22}\right) \nonumber \\
& \quad - \frac{1}{288} \frac{t v^2}{P_v^2} \left[ 3024 \widetilde{H}_{00} + 180 \widetilde{H}_{10} + 56 \widetilde{H}_{22} + 9 \widetilde{H}_{32} - 28 \eta  \left(54 \widetilde{H}_{00}+\widetilde{H}_{22}\right) \right] \nonumber \\
& \quad - \frac{1}{288} \frac{P^2 v^2}{P_v^2} \left( 56 \widetilde{H}_{20} + 27 \widetilde{H}_{30} - 196 \eta  \widetilde{H}_{20}\right) + \mathcal{O}(\text{twist-five}) \,.
\end{align}

\noindent\underline{$\bullet$~~$\widetilde{\mathfrak{E}}^{(v)}_N(t)$}
\begin{align}
\widetilde{\mathfrak{E}}_0^{(v)} & = \widetilde{E}_{00}  \,, \qquad\qquad
\widetilde{\mathfrak{E}}_1^{(v)} = \widetilde{E}_{10} - 6 \eta  \widetilde{E}_{00}  \,, \nonumber \\
\widetilde{\mathfrak{E}}_2^{(v)} & = \widetilde{E}_{20} - 10 \eta  \widetilde{E}_{10} + 4 \eta^2 \left(9 \widetilde{E}_{00}+\widetilde{E}_{22}\right)+\frac{m_N^2 v^2}{3 P_v^2}  \left(9 \widetilde{H}_{00}-4 \widetilde{H}_{22}\right) \nonumber \\
& \quad - \frac{t v^2}{4 P_v^2} \left(3 \widetilde{E}_{00} + 2 \widetilde{E}_{22}\right) - \frac{P^2 v^2}{6 P_v^2} \widetilde{E}_{20} \,, \nonumber \\
\widetilde{\mathfrak{E}}_3^{(v)} & = \widetilde{E}_{30} - 14 \eta  \widetilde{E}_{20} + 4 \eta ^2 \left(20 \widetilde{E}_{10}+\widetilde{E}_{32}\right) - 56 \eta^3 \left(4 \widetilde{E}_{00}+\widetilde{E}_{22}\right) \nonumber \\
& \quad + \frac{1}{27} \frac{v^2 m_N^2}{P_v^2} \left[ 125 \widetilde{H}_{10}-27 \widetilde{H}_{32} + 126 \eta \left(4 \widetilde{H}_{22}-9 \widetilde{H}_{00}\right) \right] \nonumber \\
& \quad - \frac{1}{216} \frac{t v^2}{P_v^2} \left[  290 \widetilde{E}_{10} + 81 \widetilde{E}_{32} - 756 \eta  \left(3 \widetilde{E}_{00} + 2 \widetilde{E}_{22}\right) \right] \nonumber \\
& \quad + \frac{1}{24} \frac{P^2 v^2}{P_v^2} \left( 56 \eta  \widetilde{E}_{20} - 9 \widetilde{E}_{30}\right) \,.
\end{align}

\noindent\underline{$\bullet$~~$\widetilde{\mathfrak{E}}^{(P)}_N(t)$}
\begin{align}
\widetilde{\mathfrak{E}}_0^{(P)} & =\widetilde{E}_{00} \,, \qquad\qquad
\widetilde{\mathfrak{E}}_1^{(P)} = \widetilde{E}_{10}-6 \eta  \widetilde{E}_{00} \,, \nonumber \\
\widetilde{\mathfrak{E}}_2^{(P)} & = \widetilde{E}_{20} - 10 \eta  \widetilde{E}_{10} + 4 \eta^2 \left(9 \widetilde{E}_{00}+\widetilde{E}_{22}\right)  \nonumber \\
& \quad + \frac{2}{9} \frac{v^2m_N^2}{P_v^2} \left(27 \widetilde{H}_{00}-2 \widetilde{H}_{22}\right) - \frac{t v^2}{6 P_v^2} \left(9 \widetilde{E}_{00}+\widetilde{E}_{22}\right) - \frac{P^2 v^2}{18 P_v^2} \widetilde{E}_{20} \,, \nonumber \\
\widetilde{\mathfrak{E}}_3^{(P)} & = \widetilde{E}_{30}-14 \eta  \widetilde{E}_{20} + 4 \eta^2 \left(20 \widetilde{E}_{10}+\widetilde{E}_{32}\right) - 56 \eta^3 \left(4 \widetilde{E}_{00}+\widetilde{E}_{22}\right) \nonumber \\
& \quad + \frac{1}{54} \frac{v^2 m_N^2}{P_v^2} \left[ 370 \widetilde{H}_{10}-27 \widetilde{H}_{32} + 168 \eta \left(2 \widetilde{H}_{22}-27 \widetilde{H}_{00}\right) \right]  \nonumber \\
& \quad + \frac{1}{432}\frac{t v^2}{P_v^2} \left[ 1008 \eta  \left(9 \widetilde{E}_{00}+\widetilde{E}_{22}\right) - 1060 \widetilde{E}_{10} - 81 \widetilde{E}_{32} \right]  \nonumber \\
& \quad + \frac{1}{144} \frac{P^2 v^2}{P_v^2} \left(112 \eta  \widetilde{E}_{20} - 27\widetilde{E}_{30}\right) \,.
\end{align}

\noindent\underline{$\bullet$~~$\widetilde{\mathfrak{E}}^{(\Delta)}_N(t)$}
\begin{align}
\widetilde{\mathfrak{E}}_0^{(\Delta )} & = -\frac{\widetilde{E}_{00}}{2 \eta } \,, \qquad\qquad
\widetilde{\mathfrak{E}}_1^{(\Delta )} = 3 \widetilde{E}_{00} - \frac{1}{4 \eta}\widetilde{E}_{10} \,, \nonumber \\
\widetilde{\mathfrak{E}}_2^{(\Delta )} & =  - \frac{1}{6 \eta} \widetilde{E}_{20} + \frac{5}{2} \widetilde{E}_{10} - 2 \eta  \left(9 \widetilde{E}_{00} + \widetilde{E}_{22}\right) + \frac{2}{9}\frac{m_N^2 v^2}{\eta  P_v^2} \left(9 \widetilde{H}_{00}+\widetilde{H}_{22}\right)  \nonumber \\
& \quad + \frac{1}{12} \frac{t v^2}{\eta  P_v^2} \left(9 \widetilde{E}_{00}+\widetilde{E}_{22}\right) + \frac{1}{36} \frac{P^2 v^2}{\eta  P_v^2} \widetilde{E}_{20} \,, \nonumber \\
\widetilde{\mathfrak{E}}_3^{(\Delta )} & =  - \frac{1}{8 \eta} \widetilde{E}_{30} + \frac{7}{3} \widetilde{E}_{20} - \frac{1}{6} \eta  \left( 110 \widetilde{E}_{10}+9 \widetilde{E}_{32}\right) + 28 \eta^2 \left(4 \widetilde{E}_{00}+\widetilde{E}_{22}\right)  \nonumber \\
& \quad + \frac{1}{216}\frac{v^2 m_N^2}{\eta P_v^2} \left[ 27 \widetilde{H}_{32}-790 \widetilde{H}_{10} - 84 \eta\left(54 \widetilde{H}_{00}+11 \widetilde{H}_{22}\right) \right]  \nonumber \\
& \quad + \frac{1}{1728} \frac{t v^2}{\eta P_v^2} \left[ 81 \widetilde{E}_{32} - 1040 \widetilde{E}_{10} - 2016 \eta \left(9 \widetilde{E}_{00}+\widetilde{E}_{22}\right) \right] \nonumber \\
& \quad + \frac{1}{576} \frac{P^2 v^2}{\eta P_v^2} \left( 27 \widetilde{E}_{30} - 392 \eta \widetilde{E}_{20} \right) + \mathcal{O}(\text{twist-five}) \,.
\end{align}

\noindent\underline{$\bullet$~~$\mathfrak{M}^{(\gamma)}_N(t)$}
\begin{align}
\mathfrak{M}_{0}^{(\gamma)} & = 0 \,, \qquad\qquad
\mathfrak{M}_{1}^{(\gamma)} =3 M_{00} \,, \qquad\qquad
\mathfrak{M}_{2}^{(\gamma)} =\frac{5}{3} M_{10} - 30 \eta  M_{00} \,, \nonumber \\
\mathfrak{M}_{3}^{(\gamma)} & =\frac{7}{6} \left[ M_{20} - 20 \eta  M_{10} + 4 \eta^2 \left(54 M_{00}+M_{22}\right) \right] \nonumber \\
& \quad - \frac{7}{36} \frac{v^2}{P_v^2} \left[ t \left(54 M_{00}+M_{22}\right)+P^2 M_{20} \right] + \mathcal{O}(\text{twist-five}) \,.
\end{align}

\noindent\underline{$\bullet$~~$\mathfrak{M}^{(\tilde{\Delta})}_N(t)$}
\begin{align}
\mathfrak{M}_{0}^{(\tilde{\Delta})} & = 0 \,, \qquad\qquad
\mathfrak{M}_{1}^{(\tilde{\Delta})} = 0 \,, \qquad\qquad 
\mathfrak{M}_{2}^{(\tilde{\Delta})} = -\frac{5 M_{10}}{6}  \,, \nonumber \\
\mathfrak{M}_{3}^{(\tilde{\Delta})} & = \frac{7}{6} \left(10 \eta  M_{10}-M_{20}\right) \,.
\end{align}

\noindent\underline{$\bullet$~~$\mathfrak{E}^{(\tilde{\Delta})}_N(t)$}
\begin{align}
\mathfrak{E}_{1}^{(\tilde{\Delta})} & = 0 \,, \qquad \qquad \mathfrak{E}_{1}^{(\tilde{\Delta})} = -\frac{3}{2 \eta } E_{00} \,, \qquad\qquad
\mathfrak{E}_{2}^{(\tilde{\Delta})} = 15 E_{00}-\frac{5}{3 \eta}  E_{10} \,, \nonumber \\
\mathfrak{E}_{3}^{(\tilde{\Delta})} & = - \frac{7}{12 \eta} \left[ 3 E_{20} -40 \eta  E_{10} + 4 \eta ^2 \left(54 E_{00}+E_{22}\right) \right] \nonumber \\
& \quad  - \frac{7}{72 \eta}  \frac{v^2}{P_v^2} \left[ t \left(54 E_{00}+E_{22}\right)+3 P^2 E_{20}-2 m_N^2 M_{20}\right] + \mathcal{O}(\text{twist-five}) \,.
\end{align}


\addcontentsline{toc}{section}{References}

\bibliography{references}

\bibliographystyle{JHEP}



\end{document}